\def\paperID{0000}
\def\confName{CVPR}
\def\confYear{2025}

\def\paperTitle{Creative Blends of Visual Concepts}

\def\authorBlock{
    Zhida Sun$^{1}$, Zhenyao Zhang$^{1}$, Yue Zhang$^{1}$, Min Lu$^1$, Dani Lischinski$^{2}$, Daniel Cohen-Or$^{3}$, Hui Huang$^{1}$\thanks{Corresponding author} \\
    $^1$Shenzhen University, $^2$Hebrew University of Jerusalem, $^3$Tel Aviv University \\
    {\tt\small \{zhdsun, a3060458966, yuezhanghci, lumin.vis, danix3d, cohenor, hhzhiyan\}@gmail.com}
}

\newif\ifreview 
\newif\ifarxiv \newcommand{\arxiv}{\arxivtrue}
\newif\ifcamera 
\newif\ifrebuttal 

\arxiv 

\pdfoutput=1
\documentclass[10pt,twocolumn,letterpaper]{article}
\ifreview \usepackage[review]{cvpr} \fi
\ifarxiv \usepackage[pagenumbers]{cvpr} \fi
\ifrebuttal \usepackage[rebuttal]{cvpr} \fi
\ifcamera \usepackage{cvpr} \fi

\usepackage{graphicx}	
\usepackage{amsmath}	
\usepackage{amssymb}	
\usepackage{booktabs}
\usepackage{times}
\usepackage{microtype}
\usepackage{epsfig}
\usepackage{caption}
\usepackage{float}
\usepackage{placeins}
\usepackage{color, colortbl}
\usepackage{stfloats}
\usepackage{enumitem}
\usepackage{tabularx}
\usepackage{xstring}
\usepackage{multirow}
\usepackage{xspace}
\usepackage{url}
\usepackage{subcaption}
\usepackage{xcolor}
\usepackage[hang,flushmargin]{footmisc}

\usepackage{array,makecell}
\usepackage{balance}
\usepackage{cuted}

\usepackage{fancyvrb}
\usepackage{fvextra}

\ifcamera \usepackage[accsupp]{axessibility} \fi

% ...

\definecolor{vermilion}{RGB}{211,84,0}
\definecolor{frenchblue}{RGB}{68,114,196}
\definecolor{pur}{HTML}{542788}
\definecolor{lem}{HTML}{b35806}
\definecolor{gre}{HTML}{01665e}
\definecolor{gol}{HTML}{8c510a}

\newcommand{\ora}[1]{{\color{vermilion}#1}}
\newcommand{\blu}[1]{{\color{frenchblue}#1}}

\newcommand{\pur}[1]{{\color{pur}#1}}
\newcommand{\lem}[1]{{\color{lem}#1}}
\newcommand{\gre}[1]{{\color{gre}#1}}
\newcommand{\gol}[1]{{\color{gol}#1}}

\newcommand{\sysname}{Creative Blends}

\ifarxiv  \fi

\newcommand{\R}[1]{{%
    \textbf{%
        \ifstrequal{#1}{1}{\textcolor{red}{R#1}}{%
        \ifstrequal{#1}{2}{\textcolor{blue}{R#1}}{%
        \ifstrequal{#1}{3}{\textcolor{magenta}{R#1}}{%
        \ifstrequal{#1}{4}{\textcolor{teal}{R#1}}{%
                           \textcolor{cyan}{R#1}%
        }}}}%
    }%
}}

\usepackage{xr-hyper}

\makeatletter
\newcommand*{\addFileDependency}[1]{
  \typeout{(#1)}
  \@addtofilelist{#1}
  \IfFileExists{#1}{}{\typeout{No file #1.}}
}

\makeatother

\definecolor{cvprblue}{rgb}{0.21,0.49,0.74}
\usepackage[pagebackref,breaklinks,colorlinks,allcolors=cvprblue]{hyperref}
\usepackage[capitalize]{cleveref}
\crefname{section}{Sec.}{Secs.}
\crefname{table}{Table}{Tables}
\crefname{figure}{Fig.}{Figs.}

\ifarxiv \crefname{appendix}{App.}{Apps.}
\else \crefname{appendix}{Suppl.}{Suppls.} \fi

\begin{document}
\title{\paperTitle}
\author{\authorBlock}
\maketitle

\begin{strip}
\begin{minipage}{\textwidth}\centering
\vspace{-40pt}
\includegraphics[width=\textwidth]{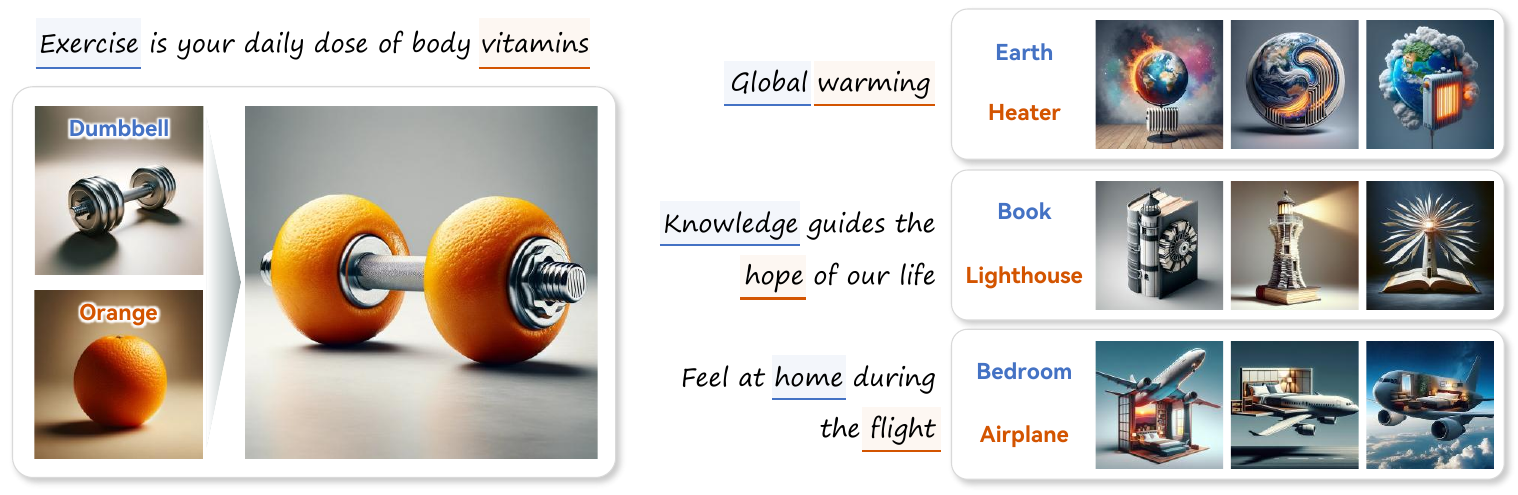}
\captionof{figure}{The \sysname\ system leverages textual user input to generate visual blends, offering a platform for creative exploration. The system broadens the spectrum of design possibilities by mapping abstract concepts -- highlighted in \blu{blue} and \ora{orange} colors -- onto tangible physical objects through constructed metaphorical associations. These mapped objects are then seamlessly integrated based on shared attributes, resulting in conceptually meaningful and visually cohesive blends.}
\label{fig:teaser}
\end{minipage}
\end{strip}
\begin{abstract}
Visual blends combine elements from two distinct visual concepts into a single, integrated image, with the goal of conveying ideas through imaginative and often thought-provoking visuals.
Communicating abstract concepts through visual blends poses a series of conceptual and technical challenges.
To address these challenges, we introduce {\normalfont \sysname}, an AI-assisted design system that leverages metaphors to visually symbolize abstract concepts by blending disparate objects. 
Our method harnesses commonsense knowledge bases and large language models to align designers' conceptual intent with expressive concrete objects. Additionally, we employ generative text-to-image techniques to blend visual elements through their overlapping attributes. 
A user study (N=24) demonstrated that our approach reduces participants' cognitive load, fosters creativity, and enhances the metaphorical richness of visual blend ideation.
We explore the potential of our method to expand visual blends to include multiple object blending and discuss the insights gained from designing with generative AI.
\end{abstract}

\section{Introduction}

Visual blending is a powerful graphic design and communication technique, offering a creative means to convey novel and groundbreaking ideas~\cite{Cunha2020LetsFT, 10.1145/3290605.3300402}.
Visual blends strategically combine two distinct objects into a unified composition, achieving a harmonious effect through unique and compelling designs~\cite{10.1145/3290605.3300402, 10.1007/s00354-020-00107-x, 10.1145/3411764.3445089}.
For instance, blending a dumbbell with an orange might symbolize the relationship between fitness and nutrition, highlighting how physical exercise complements proper nutritional intake in maintaining good health (refer to Figure~\ref{fig:teaser} on the left).
Such representations are ubiquitous in our daily lives, communicating symbolic messages from simple graphic design to intricate visual arts~\cite{10.1086/209396, benczes2009visual}.
The widespread use of visual blends stems from their capacity to engage audiences and effectively communicate complex or abstract concepts in a memorable and impactful manner.

While visual blends excel at grabbing audience attention, creating good ones presents a non-trivial design challenge.
Key design considerations include selecting appropriate objects, determining suitable blending attributes, and ensuring harmonious integration of blended elements.
A major conceptual challenge lies in identifying symbolic objects that effectively convey the intended message while preserving their individual recognizability~\cite{10.1145/3290605.3300402}.
From a technical standpoint, creating visually appealing and coherent blends requires meticulous attention to blending techniques and adherence to established design principles, making the process both skill-intensive and time-consuming.
To convey abstract concepts or experiences, visual blends often employ metaphors~\cite{MetaCLUE_10204033,grady1999blending,10.1145/3290605.3300402}, necessitating the selection of suitable source-domain objects that are both semantically relevant and visually compatible.

In the realm of Artificial Intelligence Generated Content (AIGC), the generation of images is a rapidly evolving field. 
Existing text-to-image (T2I) generation techniques, while capable of producing images from textual input, also provide an alternative tool for accelerating and simplifying the process of creating visual blends.
However, current T2I models encounter challenges when handling abstract descriptions, as they often lead to images with distorted or nonsensical textual elements~\cite{Liao_Chen_Fu_Du_He_Wang_Han_Zhang_2024}.
This limitation highlights the need for a deeper understanding of human cognition in transforming abstract content into visual elements in images.
One intriguing aspect of addressing this problem involves incorporating Conceptual Metaphor Theory (CMT) into the image-generation process~\cite{MetaCLUE_10204033,lakoff2008metaphors}.
By grounding abstract concepts or experiences in concrete objects and their interrelations, CMT posits that metaphor, beyond being a rhetorical device, offers a way to transform disparate concepts into more relatable and comprehensible information for users.
However, current research has not yet fully explored how to represent abstract concepts in image generation, particularly when employing metaphors to combine multiple objects and their relationships.

Our research pioneers the intersection of visual metaphors and image generation, offering a unique lens for understanding the cross-modal relationships underpinning creative ideation in visual blends.
We explore the potential of AIGC techniques for producing visually blended content and develop a system for semantically comparing the resulting design options.
Our key innovation integrates metaphors with commonsense reasoning at both object and attribute levels, enabling the creation of visually diverse and semantically rich blends that resonate with human metaphorical cognition.
By leveraging diverse visual-textual correspondences, we aim to unlock new creative possibilities and generate blended visuals that are not confined by predefined visual paradigms (e.g., specific shapes and styles) or literal representations. 
Inspired by Lakoff and Johnson's foundational work on metaphors~\cite{lakoff2008metaphors}, we consider language a proxy for connecting abstract ideas with visual elements, establishing a robust foundation for conceptual understanding and creative expression~\cite{10.1145/3106625}.

In line with our research, we introduce \sysname, an AI-powered system that assists users in generating visual blend ideas by incorporating metaphors derived from user input.
Informed by interviews with eight design practitioners, we identified their needs and obstacles when creating visual blends with concrete objects, especially in drafting design options for abstract concepts.
Here, ``concepts'' are keywords extracted from user-provided expressions, while ``objects'' are tangible visual elements representing these ideas.
Our method leverages metaphors and image generation techniques, complemented by large language models (LLMs) and commonsense knowledge bases, to enrich the diversity and metaphoricity of design outcomes.
A within-subject study with 24 participants compared \sysname{} to a baseline that combined ChatGPT (built with GPT-3.5 integrated with DALL·E 3) and Google Search.
Results showed that \sysname{} significantly enhances creativity, metaphoricity, and user experience in generating visually compelling ideas for abstract concepts.
Participants appreciated its ability to generate diverse outputs and inspire innovative ideas through metaphorical blends.
This research underscores the value of specialized AI tools in unlocking the creative potential of generative models, bridging gaps in user understanding of AI capabilities, and expanding the possibilities of metaphor-informed visual design.
The main contributions of this work are three-fold:
\begin{itemize}[leftmargin=*]
    \item We propose a metaphor-inspired approach to visual blending that identifies conceptually relevant objects through commonsense reasoning and combines them based on attribute similarity; 
    \item We introduce \sysname, an AI-assisted creativity support system that transforms user-provided textual input into visually blended concepts, facilitating the exploration of diverse visual-textual correspondences;
    \item Through a comprehensive evaluation, we provide insights into designing abstract concepts using metaphors within the context of AIGC, showcasing the system’s potential to advance creative workflows and expand the boundaries of visual ideation.
\end{itemize}

\section{Related Work}

This section reviews prior work on metaphorical visual design, examines AI-driven image generation for creativity support, and evaluates methods for representing abstract concepts visually.

\subsection{Metaphors in Visual Design}

Metaphors are powerful tools in visual design that engage audiences by suggesting meaning rather than explicitly stating it. 
By linking familiar concepts (source domain) to visual elements (target domain), they promote intuitive understanding and foster deeper audience engagement~\cite{4308795}.
Moreover, by leaving room for interpretation, metaphors encourage active participation, prompting users to uncover nuanced connections within the visual elements~\cite{Duit1991-DUIOTR}.
This characteristic of metaphors is utilized in visual design to facilitate data comprehension and emotional connection~\cite{da2016semantic}.
For example,  
Sun et al.~\cite{10.1145/3357236.3395475} used a postcard with a landscape painting to represent users' physical and mental health metaphorically.
\textit{Metamorpheus}~\cite{10.1145/3613904.3642410} used visual metaphors to facilitate the exploration and reflection of users' dream experiences in a meaningful way.
These representations enhance the intuitive perception and interpretation of information, while metaphors serve as visual cues to facilitate understanding of underlying meaning.

Beyond their ability to engage and inform, metaphors also excel at creative communication~\cite{doi:10.1207/s15327868ms}, a strength evidenced by their extensive use in advertising and graphic design.
Prior research has examined the structural topology of metaphors in visual representation~\cite{Peterson02012019}, identifying three primary categories: juxtaposition, fusion, and replacement structures. 
In the context of fusion-based visual metaphors, Chilton et al. pioneered \textit{VisiBlends}~\cite{10.1145/3290605.3300402} to assist novices in creating initial blend prototypes by overlaying two objects with the same shape.
They later introduced \textit{VisiFit}~\cite{10.1145/3411764.3445089}, a tool that aids novice users in collaboratively generating visual blends through brainstorming, synthesizing, and iterating. 
Other research efforts have also produced creative support tools, such as \textit{MetaMap}~\cite{10.1145/3411764.3445325}, to assist users in incorporating metaphors into their visual designs.
While existing research has advanced the creation of visual metaphors, it often overlooks the diverse interpretations that specific imagery can evoke.
This narrow focus limits the expressive potential of visual metaphors and constrains the creation of innovative representations of meaning.  
Our research aims to expand the boundaries of design possibilities by leveraging visual-textual correspondence to explore diverse interpretations of semantic concepts, enabling designers to explore a wider range of metaphorical visual designs.

\subsection{Support Image Creation with Generative AI}

As an emerging field, research is increasingly exploring the role of generative AI in creative image design processes.
Prior work has examined the potential of generative models to enhance creativity~\cite{10.1145/3491101.3503549,10.1145/3664595}, laying the groundwork for our exploration of visual blend ideation.
Several works focus on \textit{divergent challenges}, aiming to expand creative possibilities by enabling users to explore diverse outputs.
Tools such as \textit{CreativeConnect}~\cite{10.1145/3613904.3642794} promote divergent ideation by recombining visual references.
Other works like \textit{GANCollage}~\cite{10.1145/3563657.3596072} use mood boards to encourage broad exploration of visual styles, while \textit{DesignPrompt}~\cite{10.1145/3643834.3661588} integrates multi-modal inputs, including text, color, and images, to expand the scope of design possibilities. 
This supports our goal of leveraging multimodal information to enhance idea generation, enabling greater creative expression and broader exploration of design concepts.

Another set of works addresses \textit{convergent challenges}, catering to the need to refine and focus creative outputs by aligning generated content with user-defined goals.
For example, \textit{PromptCharm}~\cite{10.1145/3613904.3642803} and \textit{RePrompt}~\cite{10.1145/3544548.3581402} primarily use text-based inputs, automating prompt refinement and providing real-time model explanations to fine-tune and align generative outputs with user intentions.
\textit{GenQuery}~\cite{10.1145/3613904.3642847} provides iterative refinement workflows that enable users to adapt generated content through query concretization and image modification progressively. 
These methods prioritize user control, addressing the challenge of balancing generative outputs with refinement to create tailored results.
Despite advancements in generative models for image creation, their application in multi-object blending for design exploration is limited.
Our research focuses on the ideation stage, aiming to stimulate divergent thinking through the use of image-text embeddings and commonsense knowledge, enabling an iterative exploration of a diverse range of visual possibilities driven by user input.

\subsection{Visualizing Abstract Concepts}

Abstract concepts refer to ideas or thoughts devoid of physical form or concrete qualities. 
They cannot be directly sensed and often represent intangible qualities, relationships, or processes~\cite{Borghi2022}. 
Visualizing abstract concepts presents challenges due to their intangible and often subjective nature.
Recent advancements 
leverage multimodal systems, metaphorical reasoning, and creative blending techniques to make abstract concepts more interpretable, relatable, and engaging.
Prior work, such as the framework developed by Liao et al.~\cite{Liao_Chen_Fu_Du_He_Wang_Han_Zhang_2024}, has demonstrated the potential of leveraging LLMs and T2I models to generate visual representations of abstract concepts like morality, fairness, and priority. Similarly, Liu et al. introduced \textit{Opal}~\cite{10.1145/3526113.3545621}, a multimodal framework that deals with abstract words with recognizable subjects like symbols, showing how integrating text and visual elements fosters the understanding of abstract narratives.
These works leverage T2I generation as a powerful tool for creating visual interpretations of abstract concepts.

Another body of research employs metaphors~\cite{chakrabarty-etal-2023-spy} to bridge the gap between abstract concepts and visual representations.
For instance, \textit{Vismantic}~\cite{xiao2015vismantic} emphasizes the use of semantic structures to guide the generation of meaningful blends, demonstrating how metaphorical visuals rooted in commonsense knowledge can enhance the interpretation of abstract ideas.
Cunha et al.~\cite{10.1007/s00354-020-00107-x} demonstrate how blending visual elements could produce creative representations of abstract ideas, such as combining facial expressions and symbols to create new emojis that convey complex meanings.
In contrast to prior approaches that focus on specific forms of blending or concrete representations, our method emphasizes expanding the exploration space for abstract concepts. This allows for flexible and diverse visual outputs, enriching the ideation process for designers and creators.

Visual blending techniques~\cite{ge2021visualconceptualblendinglargescale,10.1145/3290605.3300402,10.1145/3411764.3445089} have also been extensively explored to craft visual representations of abstract ideas.
Ge and Parikh~\cite{ge2021visualconceptualblendinglargescale} advanced this field by integrating vision and language models to synthesize blended visuals, enhancing the conceptual alignment between input ideas and their outputs.
\textit{PopBlends}~\cite{10.1145/3544548.3580948} leverages LLMs to generate conceptual blends by combining semantic elements from distinct domains. 
Building on existing research, our approach seeks to broaden the scope of image generation by exploring the integration of diverse objects and creating complex blending relationships between them.
Utilizing a commonsense knowledge base, we systematically analyze the potential relationships within each combination, ensuring that the representation of these concepts is grounded in and aligns with human metaphorical reasoning.
This also enables a more diverse exploration space and a more meaningful, relatable visual representation for abstract concept ideation.

\section{Formative Study}


To understand how designers create visual blends and uncover potential challenges, we conducted in-depth interviews with eight practitioners, four from advertising and four from graphic design.
The participants included both amateur and professional designers, distinguished by whether they were engaged in full-time design-related work.
We employed a semi-structured interview format to engage in open-ended discussions about a specific design task. 
Additionally, we used the think-aloud protocol, encouraging participants to verbalize their thoughts, concerns, and plans in detail throughout the design process to gain deeper insights.

\subsection{Interview Setup and Process}

We recruited eight participants (5 female, 3 male) for the study, with ages ranging from 21 to 31 years ($Mean = 24, SD = 4.04$). 
All participants had prior experience incorporating visual metaphors into their design work and held formal education in visual design. Three participants were full-time design professionals, while the remaining five were design students who integrated visual metaphors into their personal projects.
During the study, participants first shared their previous experiences designing visual metaphors and were tasked with creating visual blends focused on a specific topic (i.e., ``\textit{Exercise fuels your body like vitamins}'', a theme chosen for its relatability and clarity). 
We emphasized the design process over the final product, requesting participants to detail the evolution of their ideas.
Interview questions covered the major design activities, exploring how participants conceptualized their ideas, identified sources of inspiration, and transformed concepts into prototypes. 
Toward the end of the interviews, participants were asked to reflect on the most challenging aspects of the design process and to articulate their specific needs for AI tools that could streamline their workflow and enhance their creative output in the context of visual blend design.

\subsection{Findings}

The feedback from participants revealed that they followed three main iterative processes when designing visual blends: ideation, gathering materials, and implementation. Among them, the ideation stage was cited by most participants (6 out of 8) as the most challenging and time-consuming phase. 
One participant (E5, Female, 21) emphasized the time-consuming nature of ideation, stating, ``\textit{...ideation is a major bottleneck. I need to brainstorm broadly to find the best solution, but this process is limited by my personal knowledge and requires constant iteration}''. 
According to her, the primary difficulties stemmed from selecting appropriate visual elements, identifying their commonalities, and determining how to distribute and arrange them effectively.
To better understand the ideation process of creating visual blends and address the challenges and needs voiced by participants, we further subdivided ideation into three sub-steps: idea generation, concept visualization, and visual integration.

\subsubsection{Idea Generation}

Participants primarily concentrated on identifying key terms, conceptualizing the desired visual message, and choosing figurative objects that can effectively represent the expression.
For all participants, keyword extraction emerged as the universal initial step when presented with a design topic.
After that, they attempted to identify appropriate expansions or elaborations of these keywords.
However, several participants, including professional designers, highlighted the challenges in generating innovative design concepts, especially when dealing with ambiguous or abstract ones.
One participant (E1, Female, 31) emphasized the difficulty of initiating the creative process even for experienced professionals and reflected that ``\textit{...initial concepts immediately flashed in my mind, but they often default to clichés, lacking novelty and engagement}''.
Another participant (E4, Female, 28) echoed this sentiment and expressed a specific need for AI assistance in overcoming this hurdle. 
She envisioned an AI tool capable of generating a wide range of conceptual options as a starting point for further refinement.

\subsubsection{Concept Visualization}

Participants reported significant challenges in transforming abstract concepts into appropriate concepts related to visual representation.
Identifying concrete representations for these abstract ideas was particularly time-consuming and labor-intensive, as one participant (E3, Female, 27) highlighted.
This challenge intensified when design topics involved multiple abstract concepts, requiring considerable effort to visually translate and integrate them cohesively.
To address this, participants expressed the need for tools or methods that could systematically examine the input abstract concepts from diverse perspectives and viewpoints.
Our observations revealed that participants assessed how connected the selected concepts were when designing.
Therefore, we prioritize the development of features that facilitate semantic relevance analysis for designers.
This would enable a more comprehensive exploration and understanding of the concepts, better equipping the designer to create meaningful visual representations that capture the essence of the original abstract expressions.

Furthermore, participants universally recognized the importance of this step in bringing their ideas to fruition.
At this step, designers are required to move beyond textual descriptions and manifest concepts visually.
Two participants (E2, Female, 21; E6, Male, 21) mentioned that they would boost their creativity by taking reference for existing design solutions, as E6 reflected ``\textit{...benchmarking against existing work sparks innovation and allows for iterative refinement}''.
The visual output of a designer's ideas relies on pre-defined concepts and is also a means of optimizing their design.
However, designers encountered instances where suitable presentation materials for their envisioned design were unavailable, thereby restricting their creative potential.
To overcome this obstacle, participants expressed a need for rapid access to a diverse range of reference images.

\begin{figure*}[t]
  \centering
  \includegraphics[width=\linewidth]{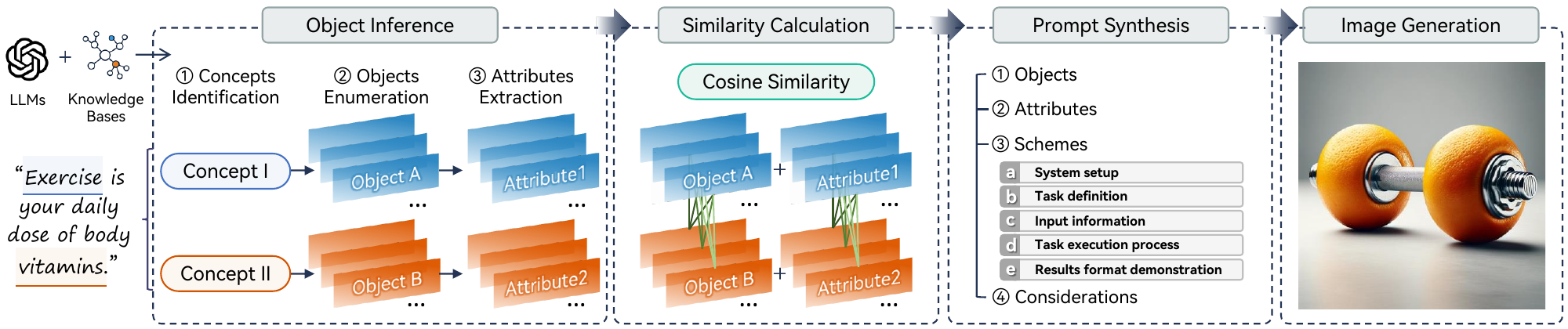}
  \caption{\sysname\ operates through a multi-stage pipeline.
  The initial stage involves concept inference to identify relevant objects and their attributes. Subsequently, a similarity-based selection process empowers users to choose suitable object and attribute combinations.
  The system then explores potential blending schemes and synthesizes corresponding prompts for the T2I model, culminating in iterative image generation based on the selected prompts to support the ideation process.}
     \label{fig:pipeline}
\end{figure*}

\subsubsection{Visual Integration}

This step involves the merging of two objects to create a cohesive and creative design. 
Participants identified the primary challenge, which is determining the optimal points of connection between the two objects. 
One participant (E8, Male, 22) commented, ``\textit{Joining two things aesthetically isn't random; it needs a thoughtful and organized process.}''.
The complexity arises from the need to effectively represent individual concepts visually while maintaining a cohesive overall visual composition.
By observing the general approaches designers use to merge two objects, we found that ensuring a seamless blend requires careful consideration of factors such as composition, color, perspective, texture, and scale.
The importance of this step in assessing a designer's overall skill set is particularly emphasized in terms of combining disparate elements. 
Moreover, the process of creating visually appealing blends often involves an iterative approach of trial and error, exploring and evaluating various options throughout the creative process.
Participants suggested expanding the design solution space through the divergent exploration of conceptual attributes followed by convergent selection based on attribute similarity to reduce cognitive load during the design process.

In summary, the creation of visual blends extends beyond the mere combination of two images. It requires creativity and a solid understanding of related concepts and their interconnected attributes to achieve effective and aesthetically pleasing visual communication.
Based on the findings from the formative study, we summarized four design requirements as follows:
\begin{enumerate}[leftmargin=*, label=\textbf{R\arabic*}]
    \item Assist with innovative idea generation while possessing clear connections to the message;
    \item Facilitate semantic relevance analysis and visual integration based on attribute similarity;
    \item Enable rapid access to diverse reference materials or design examples; 
    \item Allow users to try multiple ideas and compare them.
\end{enumerate}

\section{The \sysname\ System}

\begin{figure*}[t]
  \centering
  \includegraphics[width=\linewidth]{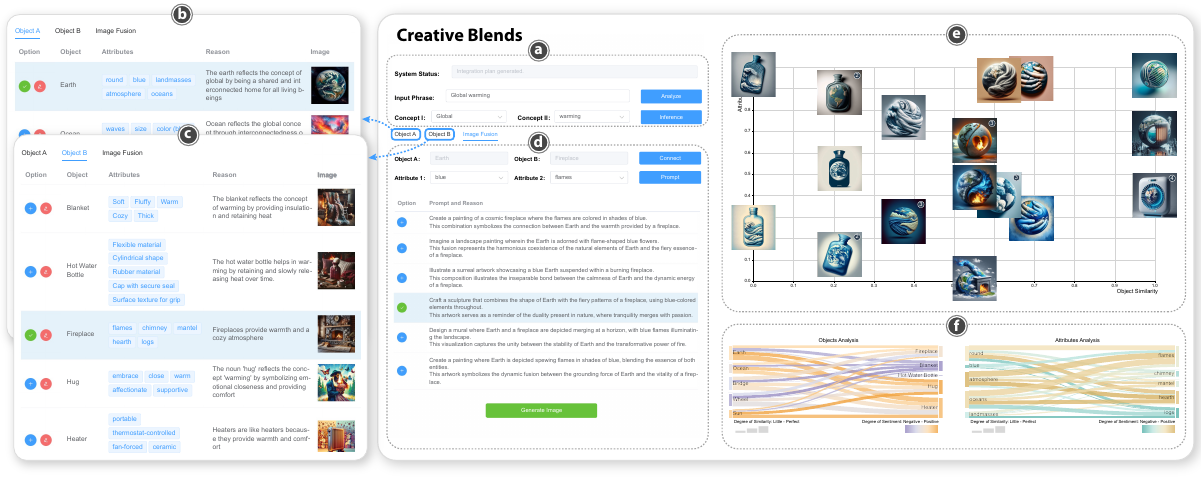}
  \caption{The user interface of \sysname\ showcases an example of generated results for ``global warming''. The interface consists of four distinct modules: the expression input module (a), the prompt exploration module (b, c, d), the visual blend exploration module (e), and the similarity visualization module (f).
  }
     \label{fig:interface}
\end{figure*}

This section introduces \sysname, an ideation support system designed to enhance designers' creativity by identifying and combining relevant objects and attributes for visual blending. Grounded in insights from prior research and the formative study, \sysname{} follows a multi-stage pipeline (Figure \ref{fig:pipeline}).
The system begins by semantically analyzing user input, using LLMs and external knowledge bases to map abstract concepts to concrete objects and attributes. 
To aid in selecting complementary elements, it provides visualized similarity and sentiment scores, facilitating informed decision-making during the ideation process.
By leveraging LLMs, \sysname{} generates detailed descriptions of potential visual blends based on user-selected inputs, blending schemes, and considerations. 
These descriptions are crafted into complete solutions, comprising prompts for the DALL·E 3 model, to enable rapid prototyping of diverse visual blend concepts. 
To support iterative creativity, users can save and revisit their outputs for refinement and comparison.
The following subsections detail the design, technical implementation, and user experience of \sysname.

\subsection{Design Goals}

Our formative study revealed key design requirements for visual blends, emphasizing the recurring challenges designers encounter in transforming abstract concepts into visual representations through the use of metaphors.
In response, we develop the \sysname\ system to support the design ideation of visual blends for a broad spectrum of abstract expressions.
The system empowers designers to enhance their creativity when searching concrete imagery and enables the design to convey the underlying meaning of the expression. 
In light of the design requirements, the design goals of \sysname\ are as follows:

\begin{enumerate}[leftmargin=*, label=\textbf{G\arabic*}]
    \item \textbf{Inferring potential objects with metaphors}. Blended imagery necessitates associated objects to symbolize abstract concepts (\textbf{R1}), with the shared attributes of these objects providing a common ground for guiding the direction of the blending process (\textbf{R1, R2}).
    \item \textbf{Exploring similarity-based exemplars within diverse options}. The visual harmony of blended images is ensured by maintaining similarities between the objects and their attributes (\textbf{R2}). The system should support the exploration of diverse object combinations and their shared attributes to inspire creative representation (\textbf{R4}).
    \item \textbf{Offering sample designs as inspiration for creative ideation}. The system should provide sample designs based on designer-selected objects and their identified commonalities, offering valuable and timely inspiration for designers (\textbf{R3}).
    \item \textbf{Iterating the exploration on potential design choices}. The system should enable designers to easily track, compare, and iterate on blended results, encouraging creative ideation with a wide range of design possibilities (\textbf{R4}).
\end{enumerate}

\subsection{System Design}

\begin{figure*}[t]
  \centering
  \includegraphics[width=0.7\linewidth]{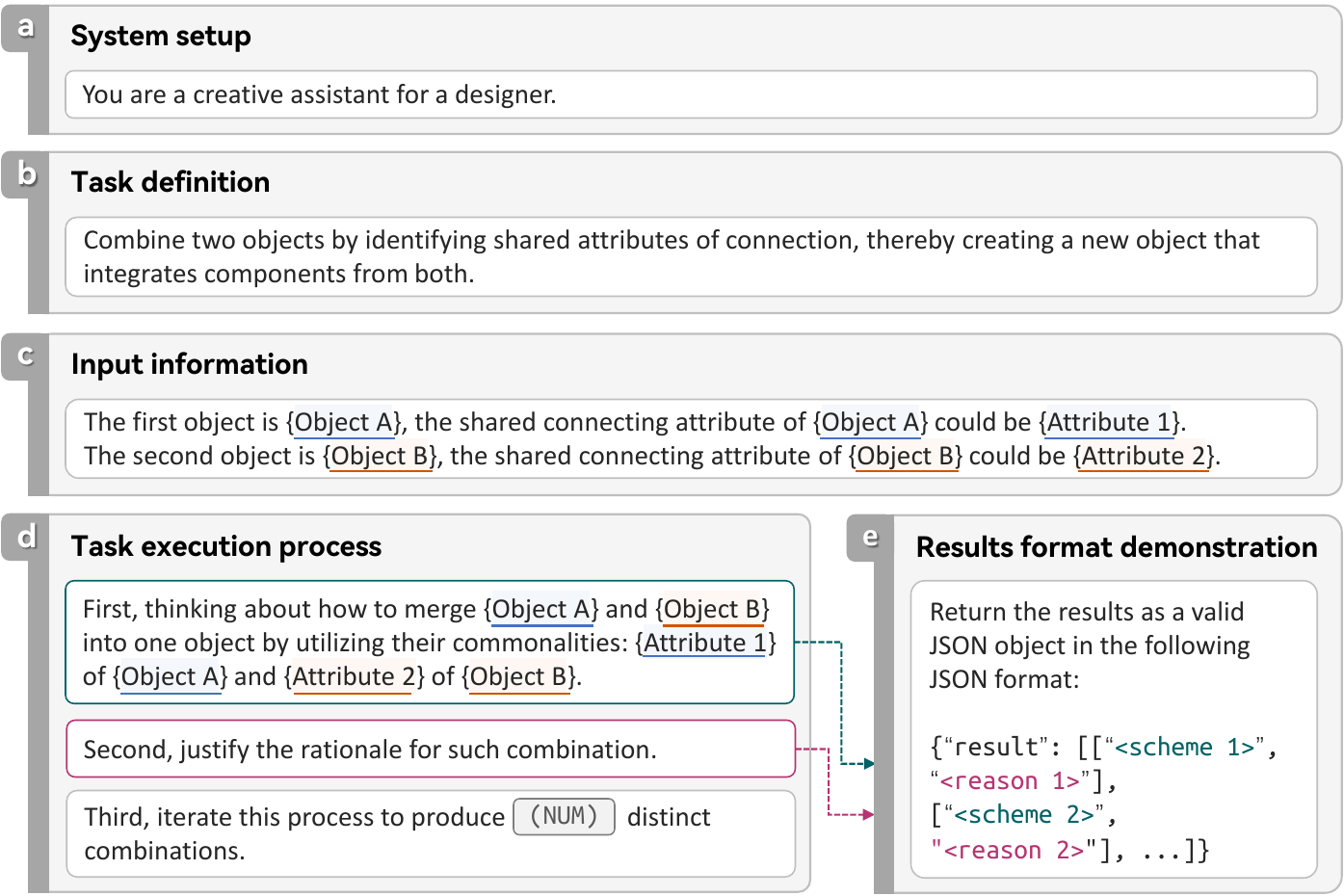}
  \caption{The scheme generation prompt is structured into five key modules: (a) system setup, (b) task definition, (c) user input, (d) task execution process, and (e) results format demonstration. The (d) task execution process module outlines the methods and rationales for considering potential blending schemes. The results, processed by the (e) results format demonstration module, are returned in a standardized JSON format, ensuring compatibility with downstream processes. 
  }
     \label{fig:prompt-template}
\end{figure*}

The \sysname\ system incorporates visualizations to assist users in exploring the generated visual blends.
Figure \ref{fig:interface} illustrates the user interface of \sysname, which processes the user input in the following manner.
The system initially decomposes the provided expression into its constituent parts of speech. Users are then prompted to select the concepts they are interested in for further exploration. Based on the selected results, the system identifies related objects that can be metaphorically linked to the chosen concepts and extracts their physical attributes (\textbf{G1}).
To facilitate the exploration of object and attribute combinations, \sysname\ employs Sankey diagrams to visualize semantic and similarity relationships.
The diagram's node-link representation, coupled with color-coded edges, enhances data clarity and readability, empowering users to explore diverse combinations and foster creative ideation (\textbf{G2}).
Additionally, the system generates tailored prompts based on the user-selected objects and attributes, providing guidance for the blending process. Other relevant factors, such as the intended expression, chosen metaphors, and applicable design restrictions, are also incorporated into these prompts, which serve as input for T2I models (\textbf{G3}).
Finally, the system generates and displays the blended results on a 2D canvas, enabling users to explore, compare, and identify the preferred ideation outcomes (\textbf{G4}).

\subsection{Implementation}

\subsubsection{Identifying Metaphorical Objects and Their Attributes}

To achieve our design goals, we initially tokenize the user input into adjectives, nouns, and verbs, allowing users to select the appropriate concepts from them (Figure~\ref{fig:interface}a).
The entire user input will later provide the model with a comprehensive context, allowing it to understand the intended concept better and identify any underlying metaphors.
Next, we integrate ConceptNet~\cite{10.5555/3298023.3298212}, an external knowledge base, with GPT-3.5-turbo~\cite{NEURIPS2020_1457c0d6} to assist users in discovering objects connected to the given concepts.
Leveraging CMT, we embed the metaphorical expression template, ``\texttt{\{Concept\} is like [a/an] \{Object\}}'', into the prompts to identify related objects (\textbf{G1}).
Specifically, ConceptNet is leveraged to extract the top 50 semantically related objects for the target concept, which are then incorporated into the prompt to guide GPT in selecting the most appropriate ones.
For each object, ConceptNet is further utilized to identify associated attributes, which are integrated into the GPT prompts as contextual references, enhancing its capacity to suggest relevant attributes.
During each query iteration, the system presents five objects to the user, enabling iterative refinement and updates until a suitable selection is achieved.
Subsequently, users can review the rationale behind the metaphorical connections between each object and its associated concept, along with the top five attributes identified for each concept-object pair (Figure~\ref{fig:interface}b\&c).
As shown in Figure~\ref{fig:teaser} (left), if the user selects ``vitamins'' as a concept, the system suggests related objects like ``orange'', ``medicine'', and ``egg''.
Upon choosing ``orange'' as the object, GPT suggests that the metaphorical connection between ``vitamins'' and ``orange'' arises from the perspective that ``it contains vitamin C, which is essential for health and well-being''. 
Additionally, the system identifies related attributes of ``orange'', such as ``round'', ``orange color'', and ``juicy''.

\subsubsection{Calculation of Similarity and Sentiment Scores}

To facilitate the exploration of various blending options, we incorporate similarity and sentiment scores. 
These scores help users identify and compare similarities among different objects or attributes and understand the overall harmony of blended results.
We leverage a pre-trained CLIP model~\cite{pmlr-v139-radford21a}, which encodes both image and text data into a shared embedding space, thereby capturing the semantic relationship between textual and visual modalities. 
By computing cosine similarity in the CLIP text embedding space~\cite{richardson2024popsphotoinspireddiffusionoperators}, we assess the similarity among objects and evaluate their likeness.
The similarities between attributes can also help designers identify commonalities among objects, which can serve as anchors during the blending process (\textbf{G2}).
The Sankey diagrams represent the similarity scores between objects or attributes using the width of their connecting links (Figure~\ref{fig:interface}f), making it easier for users to understand the relationships between different options and aiding them in observing and comparing various generation outcomes (\textbf{G4}).
To preserve the relationships among the original data values and improve the visual clarity of the similarity matrix, we employ the Min-Max normalization~\cite{10.1007/978-3-031-42536-3_33} to standardize the similarity scores.

We use the DistilBERT model~\cite{sanh2020distilbertdistilledversionbert} to perform sentiment analysis on text descriptions of objects or attributes, thereby assisting users in making informed selections and reducing the likelihood of misinterpretations due to contextual variations or other factors.
Sentiment scores are calculated using the confidence level $C$ of the sentiment label, where ``positive'' labels are equal to $C$ while ``negative'' labels are calculated as $(1-C)$.
We employ color variations to distinguish between objects and attributes while ensuring that the color temperature gradient correlates with semantic sentiment changes.
The Sankey diagrams use \pur{purple} and \lem{orange} to represent negative and positive sentiments for objects, and \gre{green} and \gol{gold} for attributes, respectively.
To determine the overall sentiment between two objects or attributes, we average their individual sentiment scores.
Subsequently, we apply quantile normalization to the overall sentiment scores to ensure that the data distributions are similar across different samples. This also makes the color distribution in the Sankey diagrams more balanced.

\begin{figure*}[t]
  \centering
  \includegraphics[width=0.7\linewidth]{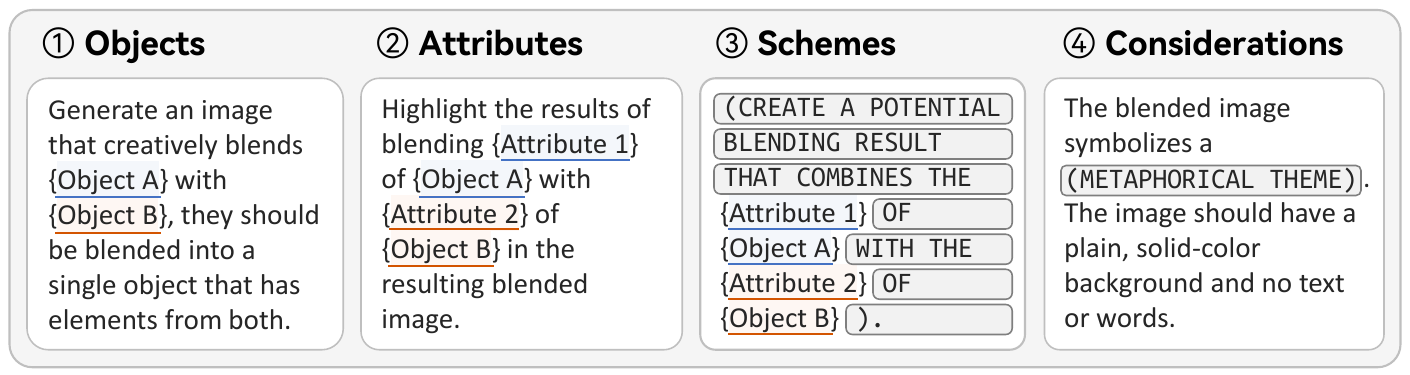}
  \caption{The final prompt is composed of four distinct modules: (1) objects, (2) attributes, (3) schemes, and (4) considerations. The schemes and metaphorical themes (marked in capital letters with a grey background), essential elements of the prompt, are dynamically generated by the GPT in response to the scheme and metaphor generation prompts. 
  }
     \label{fig:prompt}
\end{figure*}

\subsubsection{Generate Blending Prompts and Images.}

Following the user's selection of objects and attributes, \sysname\ generates blended images by first establishing blending schemes (Figure~\ref{fig:interface}d), and then converting the chosen scheme into a final prompt for image generation (\textbf{G3}).
To enhance user control and creative freedom, we adhered to prompt engineering best practices~\cite{Prompten79:online, doi:10.1080/10875301.2023.2227621} for blend scheme construction by outlining the GPT's role, the task, the user input, and a step-by-step process to guide the model (Figure~\ref{fig:prompt-template}).
Subsequently, the user's selected scheme, objects, attributes, and design considerations (e.g., metaphorical themes and design constraints) were integrated into the final prompt for visual blend generation (Figure~\ref{fig:prompt}). 
These design considerations were refined through iterative experimentation.
The complete prompt scripts used in our system are available in the supplementary material.
Finally, we use DALL·E 3 to generate images based on the prompt, swiftly transforming designers' creativity into reality.
The generated visual blends are displayed in the image exploration area (Figure~\ref{fig:interface}e). 
Users can explore these results based on object and attribute similarity (X and Y axes), uncovering new creative possibilities for visual blend creation (\textbf{G4}).
To provide a clearer overview of the generated content and support iterative exploration, we number each image group with the total number of images produced for that prompt, displayed in the upper-right corner.
Furthermore, to maximize design diversity, we avoid imposing unnecessary visual constraints on blend generation beyond the specified design considerations. 
This approach leads to a diverse range of design outcomes within the 2D exploration space.
For prompts that may be closely related, \sysname\ implements interactive zooming to adjust the scale of the visualization space and prevent image overlap.
Our system also facilitates diverse and iterative creative exploration by allowing users to alter specific objects when they conceive improved or alternative ideas (\textbf{G4}). Users can directly replace the objects in the system, prompting it to regenerate new images while discarding all prior visual information.

\begin{figure*}[t]
  \centering
  \includegraphics[width=0.88\linewidth]{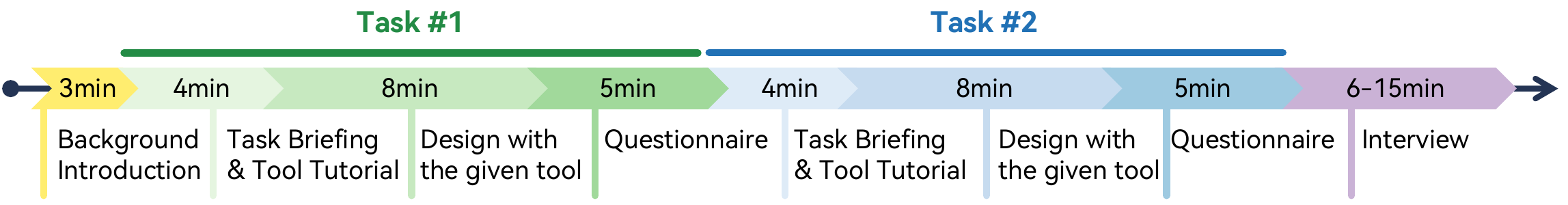}
  \caption{The user study procedure. It involved completing two tasks with both the \sysname\ and the baseline. To maintain fairness, the order of the systems and design tasks was counterbalanced.
  }
  \label{fig:study}
\end{figure*}

\subsection{User Scenario}

This section demonstrates how \sysname\ assists designers in creating visual blends. 
Alice, a graphic designer, wants to create an image that describes the concept of ``global warming'' to depict the idea of environmental protection. 
She starts by entering this phrase into the system (Figure~\ref{fig:interface}a). 
\sysname\ parses the input and identifies keywords like ``global'' and ``warming''. 
Alice selects these keywords to discover specific objects that can metaphorically represent them.
The \sysname\ system presents two lists of objects, along with their associated attributes (Figure~\ref{fig:interface}b\&c).
Two Sankey diagrams visually illustrate the connections between these objects and attributes (Figure~\ref{fig:interface}f). 
Alice initially examines the object names, their relevant attributes, and how they align with her concept (Figure~\ref{fig:interface}b\&c).
For unfamiliar objects, she can click the ``preview'' button to view generated images for more information.
She then examines the ``Objects Analysis'' Sankey diagram, which aids in object selection by displaying similarity and sentiment scores. 
As Alice explores the Sankey diagram, clicking a link automatically selects the corresponding pair of objects in the lists (Figure~\ref{fig:interface}b\&c).
She finally picks ``earth'' and ``fireplace'' as positively oriented objects with a higher degree of similarity.

Alice then delves into the system's image-blending capabilities, beginning with an examination of the ``Attributes Analysis'' Sankey diagram for guidance.
This diagram visually represents the similarity and sentiment scores among the objects' attributes. By referencing similarity, Alice can assess the coherence of the current blending results and plan her next steps.
She experiments with various attribute combinations to adjust how the objects blend within the prompt.
Ultimately, she confirms the attributes ``round'' for ``earth'' and ``flames'' for ``fireplace''.
Once satisfied with her combinations, she generates the corresponding scheme prompts along with the previously selected objects (Figure~\ref{fig:interface}d).
The final prompt then creates a new blended image in the display area, which she can enlarge and examine (Figure~\ref{fig:interface}e). 
When multiple images are created at the same location, the system displays the total number of images in the upper right corner.
After several attempts, Alice realizes she has yet to explore certain design options for visual blending.
Inspired by her exploration, she decides to utilize the editing feature to modify an object from ``fireplace'' to ``ice cream'' (Figure~\ref{fig:interface}c). 
With further experimentation and exposure to additional sample designs of visual blends with different combinations, Alice gains new ideas that she can incorporate into her subsequent creations.

\section{Hypotheses}

Previous studies indicate that creating visual blends is a challenging task~\cite{10.1145/3290605.3300402, 10.1145/3411764.3445089}. 
However, a broader range of related examples can enhance the ideation process~\cite{ECKERT2000523}. 
Additionally, reducing the cognitive load unrelated to creative tasks can boost creativity~\cite{malycha2017enhancing}.
Given the specific application of our system and the creative practices of designers, we propose the following hypotheses relative to the baseline (described in section~\ref{sec:eval}):

\begin{enumerate}[leftmargin=*, label=\textbf{H\arabic*}]
    \item Compared to the baseline, the \sysname{} system improves the overall usability by streamlining the design ideation workflow.
    \item Compared to the baseline, the \sysname{} system is capable of decreasing users' mental demand  (\textbf{H2a}), physical exertion (\textbf{H2b}), and time pressure (\textbf{H2c}) during visual blend design ideation, while simultaneously improving their satisfaction with the design ideation outcomes (\textbf{H2d}), reducing the required effort (\textbf{H2e}), and mitigating frustration during task execution (\textbf{H2f}).
    \item Compared to the baseline, the \sysname{} system is capable of generating a greater quantity (\textbf{H3a}) and a wider variety of design ideation outcomes (\textbf{H3b}) in the same amount of time, leading to improved overall designer satisfaction (\textbf{H3c}).
    \item Compared to the baseline, the \sysname{} system enables users to explore a broader range of creative ideas more easily (\textbf{H4a}), facilitates collaboration between designers and AI (\textbf{H4b}), and enhances the enjoyment of the design process (\textbf{H4c}). Additionally, users are likely to perceive the generated ideation results as more worthwhile (\textbf{H4d}), and the system's user interface enables them to focus on the task itself (\textbf{H4e}). 
    Furthermore, users can exhibit increased expressiveness during their design ideation activities (\textbf{H4f}). 
    \item Compared to the baseline, the \sysname{} system offers a more metaphorical approach to supporting user ideation activities.
\end{enumerate}

\section{Evaluation}~\label{sec:eval}

To test our hypotheses, we conducted a controlled experiment with 24 participants using a within-subject design to minimize individual differences. 
Figure~\ref{fig:study} depicts the complete evaluation process for each participant.
To ensure a fair comparison and avoid the influence of varying image generation speed and style from different models, we opted for a more rigorous approach by using ChatGPT (GPT-3.5 with DALL·E 3) and Google Search as the baseline. 
We focused on their core features, such as information search, image generation, and browsing history.
Participants undertook design ideation tasks across the \sysname{} and the baseline. 
Clear instructions were provided at the outset of each task to enable participants to explore and interact with both conditions freely.
After completing each task, we administered questionnaires to gather participants' evaluations of their experiences with the respective system conditions. 
Our questionnaire design was informed by relevant literature~\cite{HART1988139, 10.1145/2617588, brooke1996sus} and guided by our research hypotheses, which shaped the evaluation around the following dimensions: we employed the System Usability Scale (SUS) and NASA-Task Load Index (NASA-TLX) to gauge the system's usability and the cognitive load imposed on users. Additionally, we incorporated the Creative Support Index (CSI) and items focused on outcome satisfaction and metaphoricity to measure the extent to which the system fostered creative ideation, user satisfaction, and the integration of metaphor within the design process.

\subsection{Participants}

We enlisted a total of 24 participants (12 female, 12 male) aged 20 to 31 ($Mean = 24.08, SD = 3.31$) through online advertising and word-of-mouth. The participants were local university students and faculty members, with equal representation from both computer science (6 female, 6 male) and design (6 female, 6 male) backgrounds. Our selection criteria included a demonstrated interest in visual design and some prior experience in design activities~\cite{10.1145/3411764.3445325}. Additionally, all participants possessed normal color vision and had experience with conversational AI and T2I models.

\subsection{Tasks and Procedure}

During the experiment, each participant was tasked with producing visual blending ideas for the two abstract expressions: ``Smoking is like a warm welcome to death'' (T1) and ``Knowledge guides the hope of our life'' (T2). 
Based on input from professional designers, our criteria for selecting these two topics included the presence of at least two concepts, their abstract nature that allows visualization with specific objects and attributes, the potential for considerable exploration, and the inclusion of associated imagery that is either positive or negative.
Participants were free to explore their creativity by creating visual blends related to the given topics.
To maintain experimental balance, we employed a Latin Square design, yielding four possible sequences: (a) T1 (Baseline) - T2 (\sysname), (b) T2 (Baseline) - T1 (\sysname), (c) T1 (\sysname) - T2 (Baseline), and (d) T2 (\sysname) - T1 (Baseline).

\begin{figure}[t]
  \centering
  \includegraphics[width=\columnwidth]{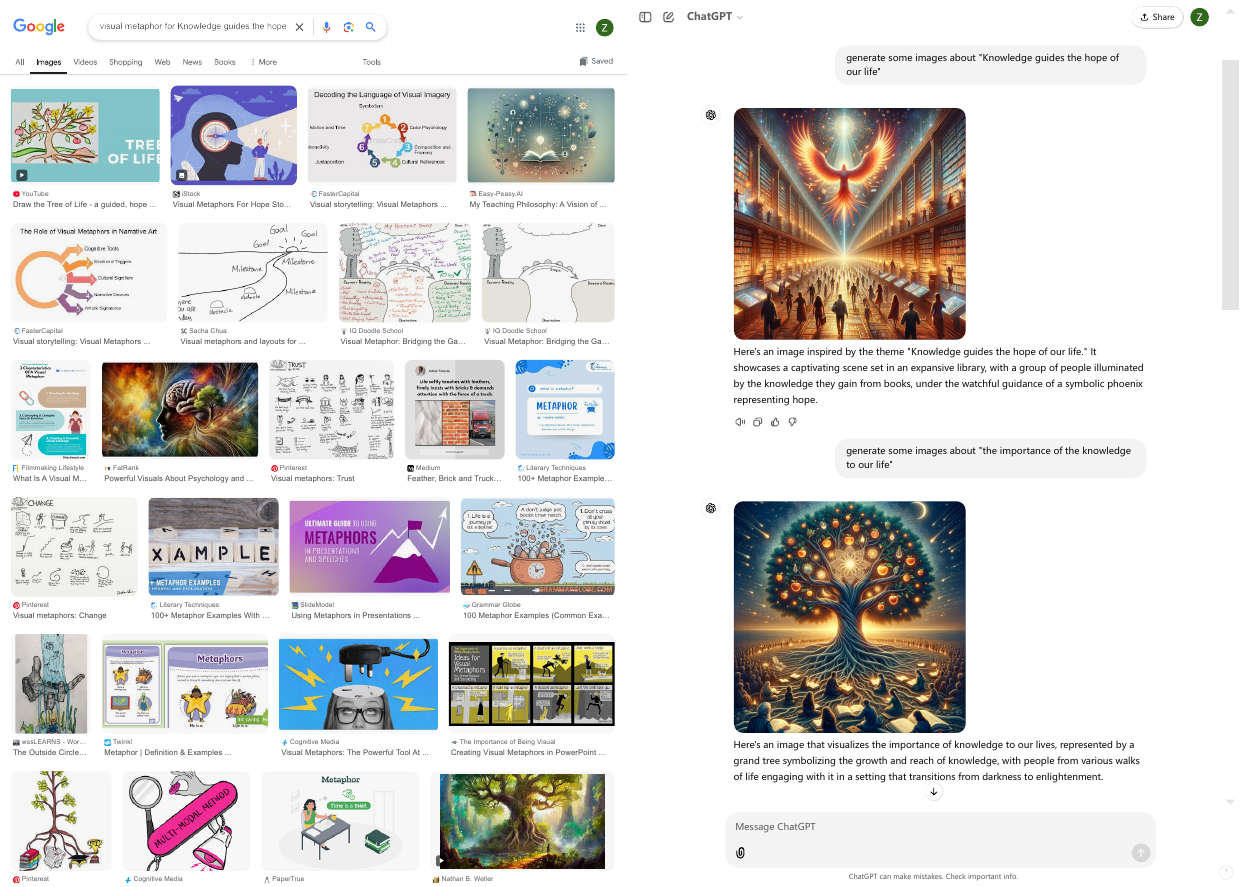}
  \caption{The baseline interface integrated Google Search and ChatGPT for both text and visual search queries.
  }
  \label{fig:baseline}
\end{figure}

Prior to the main study, we spent three minutes explaining the basic concepts of visual blends, encouraging participants to incorporate visual metaphors into their creative process.
The main study had two parts, each using either the baseline or \sysname\ system. 
The baseline interface integrated Google Search and ChatGPT into a unified side-by-side display (Figure~\ref{fig:baseline}), enabling users to interact with both platforms simultaneously.
Each task began with a 4-minute tutorial, offering an overview of the design process and step-by-step instructions, followed by a Q\&A session for clarification. 
We mitigated order effects by rotating task sequences using a Latin square design, and provided users with sample prompt words as a reference when introducing the baseline condition to ensure a fair evaluation.
After the tutorials, participants completed design tasks using the baseline or \sysname\ system within 8 minutes.
Once finished, they filled out 5-minute questionnaires about their design experience.
We also conducted 6- to 15-minute interviews to gain a deeper understanding of their experiences.
The experiment ended after the interviews, and the total duration did not exceed 60 minutes.

\begin{table*}[t]
\centering
\caption{The statistical results of user feedback with the \sysname\ and the baseline (i.e., control group), where the p-values ($-$: $p>.100$, $+$: $.050<p<.100$, $*$: $p<.050$, $**$: $p<.010$, $***$: $p<.001$) is reported. 
}
\small
\begin{tabular}{llrrrrrrrlr}
\toprule
\multicolumn{1}{c}{\multirow{2}{*}{\textbf{Category}}} &
\multicolumn{1}{c}{\multirow{2}{*}{\textbf{Factor}}}   &
\multicolumn{2}{c}{\sysname} & 
\multicolumn{2}{c}{Baseline} & 
\multicolumn{4}{c}{Statistics} & 
\multicolumn{1}{c}{\multirow{2}{*}{\textbf{Hypotheses}}} \\
\cmidrule(lr){3-4}\cmidrule(lr){5-6}\cmidrule(lr){7-10}
         &                   & Mean            & SD    & Mean     & SD      & $t$                & $Z$                & $p$                & Sig. &  \\
\toprule
\textbf{Usability}~\cite{brooke1996sus} & \textbackslash{}        & 75.000              & 9.918 & 61.458   & 21.781             & 2.547           & \textbackslash{} & 0.018            & $*$    &  \textbf{H1} \textbf{Accepted}          \\
\midrule
\multicolumn{1}{l}{\multirow{6}{*}{\parbox{1.3cm}{\textbf{Cognitive Load}~\cite{HART1988139}}}}  &
Mental Demand                  & 2.542           & 1.062 & 4.458    & 1.587              & -4.861            & \textbackslash{} & \textless{}0.001 & $***$  &   \textbf{H2a} \textbf{Accepted}         \\
                 & Physical Demand               & 1.625           & 0.824 & 2.958    & 1.967    & \textbackslash{} & -2.865           & 0.004            & $**$   &   \textbf{H2b} \textbf{Accepted}         \\
                 & Temporal Demand               & 3.250            & 1.359 & 3.833    & 1.685              & \textbackslash{} & -1.517           & 0.129            & $-$    &  \textbf{H2c} Rejected          \\
                 & Performance          & 4.583           & 0.929 & 4.333    & 1.736             & 0.598           & \textbackslash{} & 0.556            & $-$    &  \textbf{H2d} Rejected          \\
                 & Effort                  & 2.583           & 1.349 & 3.917    & 1.666              & -2.734            & \textbackslash{} & 0.012            & $*$    &  \textbf{H2e} \textbf{Accepted}          \\
                 & Frustration             & 2.625           & 1.663 & 3.625    & 1.663              & -1.931            & \textbackslash{} & 0.066            & $+$    &  \textbf{H2f} Rejected          \\
\midrule
\multicolumn{1}{l}{\multirow{3}{*}{\parbox{1.6cm}{\textbf{Outcome\\Satisfaction}}}}
        & Amount                  & 5.080            & 1.412 & 4.540     & 1.444              & 1.299           & \textbackslash{} & 0.207            & $-$    &   \textbf{H3a} Rejected         \\
                 & Diversity               & 5.380            & 1.345 & 3.670     & 1.274              & 5.058           & \textbackslash{} & \textless{}0.001 & $***$  &  \textbf{H3b} \textbf{Accepted}          \\
                 & Overall                 & 5.080            & 1.283 & 3.960     & 1.367              & 2.609           & \textbackslash{} & 0.016            & $*$    &  \textbf{H3c} \textbf{Accepted}          \\
\midrule
\multicolumn{1}{l}{\multirow{6}{*}{\textbf{Creativity}~\cite{10.1145/2617588}}}       & Exploration             & 5.708           & 1.160  & 4.083    & 1.886              & 2.985           & \textbackslash{} & 0.007            & $**$   &   \textbf{H4a} \textbf{Accepted}         \\
                 & Collaboration           & 5.500             & 1.560  & 3.667    & 2.014              & 3.350            & \textbackslash{} & 0.003            & $**$   &   \textbf{H4b} \textbf{Accepted}         \\
                 & Enjoyment              & 5.833           & 1.435 & 4.042    & 1.756              & 3.558           & \textbackslash{} & 0.002            & $**$   &   \textbf{H4c} \textbf{Accepted}         \\
                 & Results Worth Effort & 5.458           & 1.444 & 4.167    & 1.633              & 2.806           & \textbackslash{} & 0.010             & $*$    &  \textbf{H4d} \textbf{Accepted}          \\
                 & Immersion       & 5.167           & 1.903 & 5.042    & 1.706               & 0.213           & \textbackslash{} & 0.833            & $-$    &   \textbf{H4e} Rejected         \\
                 & Expressiveness          & 5.167           & 1.167 & 4.375    & 1.689               & 1.807           & \textbackslash{} & 0.084            & $+$    &   \textbf{H4f} Rejected         \\
\midrule
\textbf{Metaphoricity}    & \textbackslash{}        & 5.655           & 0.974 & 5.047    & 1.193              & 2.276           & \textbackslash{} & 0.032            & $*$    &  \textbf{H5} \textbf{Accepted}
\\
\bottomrule
\end{tabular}
\label{tbl:result}
\end{table*}

\subsection{Results Analysis}

This section presents a statistical analysis of user ratings gathered during the experimental process.
The findings, contextualized by user feedback, are presented in relation to each evaluation criterion.

\subsubsection{System Usability}

To assess system usability, we adopted the System Usability Scale (SUS)~\cite{brooke1996sus} as our questionnaire.
The means and standard deviations of the participants' ratings are presented in Table~\ref{tbl:result}.
From the participants' feedback, a Shapiro-Wilk test is conducted to examine the distribution's normality, and no evidence of non-normality is observed in the overall distribution ($W = 0.925, p = 0.075$).
A paired samples t-test comparing \sysname\ to the baseline shows significantly higher usability for \sysname\ ($t(23) = -2.547, p = 0.018$).
The effect size of Cohen's \textit{d} ($d = 0.800$) suggests a large effect on the improvement in \sysname\ usability.
Therefore, \textbf{H1} is accepted.
Two participants (P9, Male, 21; P17, Male, 21) praised \sysname\ for its simple and user-friendly interface, describing it as ``\textit{refreshing}''.
Almost all participants (22/24) found \sysname\ relatively easy to learn.
As one participant (P18, Female, 23) noted, ``\textit{While I needed some help at first to use \sysname, I was able to use it independently after practicing once.}''
Unlike the conversational interface of the baseline, \sysname\ presents generated results via a 2D mood board-like interface (Figure~\ref{fig:interface}e), providing a comprehensive overview of the output's scope.
The sentiment analysis visualizations also empower users to evaluate and refine the blending process effectively.

\subsubsection{Cognitive Load}

To evaluate cognitive load, we referred to the NASA-Task Load Index (TLX)~\cite{HART1988139}. 
The questionnaire is tailored to encompass six factors: 
mental demand (the amount of mental effort required to complete a task), 
physical demand (the degree of physical exertion needed to complete the task), 
temporal demand (the time taken to complete the task), 
performance (the impact of task completion), 
effort (the amount of effort necessary to complete the task), and frustration (the dissatisfaction experienced while completing a task).
Table~\ref{tbl:result} provides the mean and standard deviation of the participants' ratings.
The Shapiro-Wilk test reveals no evidence of non-normality in the distribution of mental demand ($W = 0.594, p =0.967$), performance ($W = 0.948, p =0.246$), effort ($W = 0.962, p =0.479$), and frustration ($W = 0.948, p =0.243$) factors. Therefore, paired samples t-tests were employed to compare the differences in these four dimensions between the two systems, as shown in Table~\ref{tbl:result}. 
However, the distributions of physical demand ($W = 0.811, p <0.001$) and temporal demand ($W = 0.870, p = 0.005$) factors deviate from normality. Consequently, the differences between the two systems in these dimensions were compared using the Wilcoxon signed-rank test.

The test results indicate that the cognitive load in the dimensions of mental demand ($t(23) = 4.861, p<0.001$), physical demand ($Z = -2.865, p = 0.004$), and effort ($t(23) = 2.734, p = 0.012$) when using \sysname\ is significantly lower than that of the baseline. 
The effect sizes of Cohen's $d$ and Pearson's correlation coefficient $r$ indicate that the reduction in mental demand ($d = -1.419$) was large, physical demand ($r = -0.414$) was moderate, and effort ($d = -0.880$) was large.
Consequently, hypotheses \textbf{H2a}, \textbf{H2b}, and \textbf{H2e} are accepted.
However, no significant differences were found in the dimensions of temporal demand, performance, and frustration, leading to the rejection of hypotheses \textbf{H2c}, \textbf{H2d}, and \textbf{H2f}.
The results show that there is little difference in the time spent and performance of the generated results between the \sysname\ and the baseline, as they both use the same underlying technologies. 
However, our approach significantly reduces the mental and physical burden on users and requires less effort from them. 
Four participants (P1, Female, 31; P2, Male, 23; P14, Male, 22; P23, Female, 30) praised the pipeline and interface applied in \sysname. 
They appreciated the time-saving benefits of not having to initiate the design process from scratch, and the generated results were more controllable, making their efforts more manageable.
Other participants further emphasized the value of the Sankey diagrams in streamlining the selection of appropriate object and attribute pairings, highlighting their ability to provide an overview and thereby minimize the cognitive effort required for individual pair evaluations.

\subsubsection{Outcome Satisfaction}

To measure outcome satisfaction, we designed the questionnaire inspired by MetaMap~\cite{10.1145/3411764.3445325}. 
The questionnaire includes three factors, i.e., satisfaction with the amount, diversity, and overall quality of the system output.
We gathered participants' ratings and conducted a Shapiro-Wilk test. The results show no evidence of non-normality in the distribution of outcome satisfaction ratings for the amount ($W = 0.939, p =0.154$), diversity ($W = 0.931, p =0.101$), and overall quality ($W = 0.936, p =0.129$). 
Therefore, paired samples t-tests were employed to compare the differences in outcome satisfaction between the two systems.
Table~\ref{tbl:result} provides the mean and standard deviation of the relevant factors.

Overall, participants were significantly more satisfied with the outcomes from \sysname\ than with those from the baseline ($t(23) = -2.609, p = 0.016$).
This large effect size (Cohen's $d = 0.845$) indicates a meaningful improvement in satisfaction with \sysname. 
Specifically, participants were significantly more satisfied with the diversity of outcomes generated by \sysname\ ($t(23) = -5.058, p = <0.001; d = 1.305$). 
However, there was no significant difference between the two systems in the number of outcomes generated. 
Therefore, hypotheses \textbf{H3b} and \textbf{H3c} are accepted, while hypothesis \textbf{H3a} is rejected.

Participants indicated that the \sysname\ system 
facilitated the generation of more varied, predictable, and actionable expressions of their ideas. 
Nearly half of the participants (11 out of 24) mentioned that the baseline often produced outcomes that deviated from their intended concepts, while \sysname\ guided users in expanding their design choices systematically. 
Some participants expressed frustration with the baseline, stating, ``\textit{I feel like my prompts are pretty clear, but I am struggling to get good results...It is really frustrating that these results are not even close to what I am looking for}'' (P8, Female, 21). 
In contrast, P17 (Male, 21) commented, ``\textit{I find it much easier to achieve the desired effect with \sysname. You can start with one idea and branch off into lots of different directions}''.
The association-based thought expansion approach for visual blend design, which links objects and their attributes, has demonstrated its effectiveness in rapidly generating diverse results.

\subsubsection{Creativity}

We borrowed the Creative Support Index (CSI)~\cite{10.1145/2617588} to assess creativity. 
We modified the CSI questionnaire to include six metrics: 
exploration (the extent to which the system supports user exploration), 
collaboration (the manner in which users collaborate with AI), 
enjoyment (the level of enjoyment participants experienced during the activity), 
results worth effort (the value derived from the effort), 
immersion (the degree of focus within the system), and 
expressiveness (the expressiveness and creativity exhibited during the activity).
Table~\ref{tbl:result} provides the mean and standard deviation of these six factors.
The Shapiro-Wilk tests indicate that the distributions of scores for all six factors are normal. 
For each factor, the following values were obtained: exploration ($W = 0.944, p = 0.205$), collaboration ($W = 0.964, p = 0.524$), enjoyment ($W = 0.946, p = 0.225$), results worth effort ($W = 0.946, p = 0.227$), immersion ($W = 0.975, p = 0.791$), and expressiveness ($W = 0.953, p = 0.307$).

The results in Table~\ref{tbl:result} show that \sysname\ significantly outperforms the baseline in supporting creativity in exploration ($t(23) = -2.985, p = 0.007$), collaboration ($t(23) = -3.350, p = 0.003$), enjoyment ($t(23) = -3.558, p = 0.002$), and results worth effort ($t(23) = -2.806, p = 0.010$). 
The substantial advantages of \sysname\ are reflected in the large effect sizes, with Cohen's d values for exploration ($d = 1.038$), collaboration ($d = 1.018$), enjoyment ($d = 1.117$), and results worth effort ($d = 0.838$).
Compared to the baseline, participants found \sysname\ to be more supportive of exploring, collaborating, and enjoying the creative process. Additionally, participants perceived a significant improvement in the value of their efforts when using \sysname.
Consequently, hypotheses \textbf{H4a}, \textbf{H4b}, \textbf{H4c}, and \textbf{H4d} are supported, whereas hypotheses \textbf{H4e} and \textbf{H4f} are refuted.

Feedback from participants suggests that our system significantly improves the exploration of design options, enhances engagement, and reduces the burden of ideation during the creative process. 
Five participants emphasized that \sysname\ helped them express abstract concepts in visual representation more effectively than the baseline (P3, Male, 27; P11, Male, 27; P13, Female, 25; P19, Male, 22; P22, Male, 26).
As P19 (Male, 22) reflected, ``...\textit{(\sysname) could really help me think outside the box and come up with some new solutions that I would not have thought of on my own}''.
\sysname\ leverages metaphors to facilitate inter-domain concept transfer. By examining the underlying relationships between disparate concepts, the system produces a more multifaceted assortment of object combinations.
Additionally, our approach fosters collaboration between users and AI. 
Like P1 (Female, 31) remarked, ``\textit{...basically, I tell \sysname\ what I am looking for, and it fills in all the gaps. It is like giving AI a blueprint, and then it builds the whole thing}''.
These findings provide empirical evidence that human-AI collaboration extends beyond simple task delegation, encompassing the ability of humans to decompose complex problems into manageable subtasks for AI processing effectively.
Moreover, two participants stated that \sysname\ increased their ability and willingness to explore (P18, Female, 23; P23, Female, 30). 
Even after the design session ended, both expressed a desire to continue exploring.

\begin{figure}[t]
  \centering
  \includegraphics[width=\columnwidth]{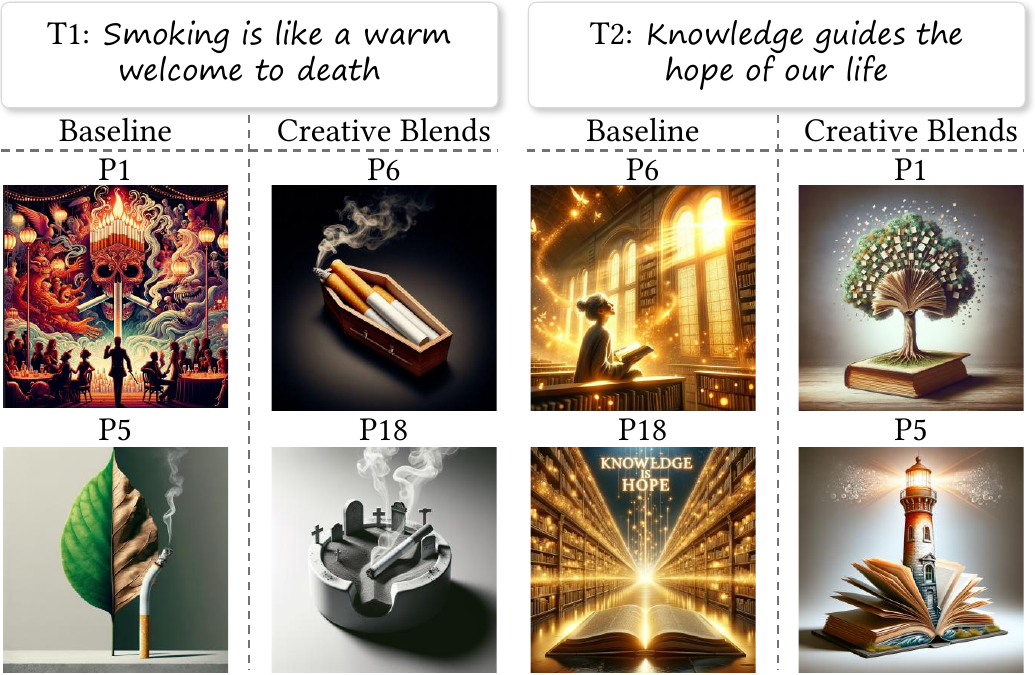}
  \caption{Eight sample outputs generated by the \sysname\ and the baseline for topics of T1 and T2.}
  \label{fig:user-study}
\end{figure}

\begin{figure*}[ht]
  \centering
  \includegraphics[width=\linewidth]{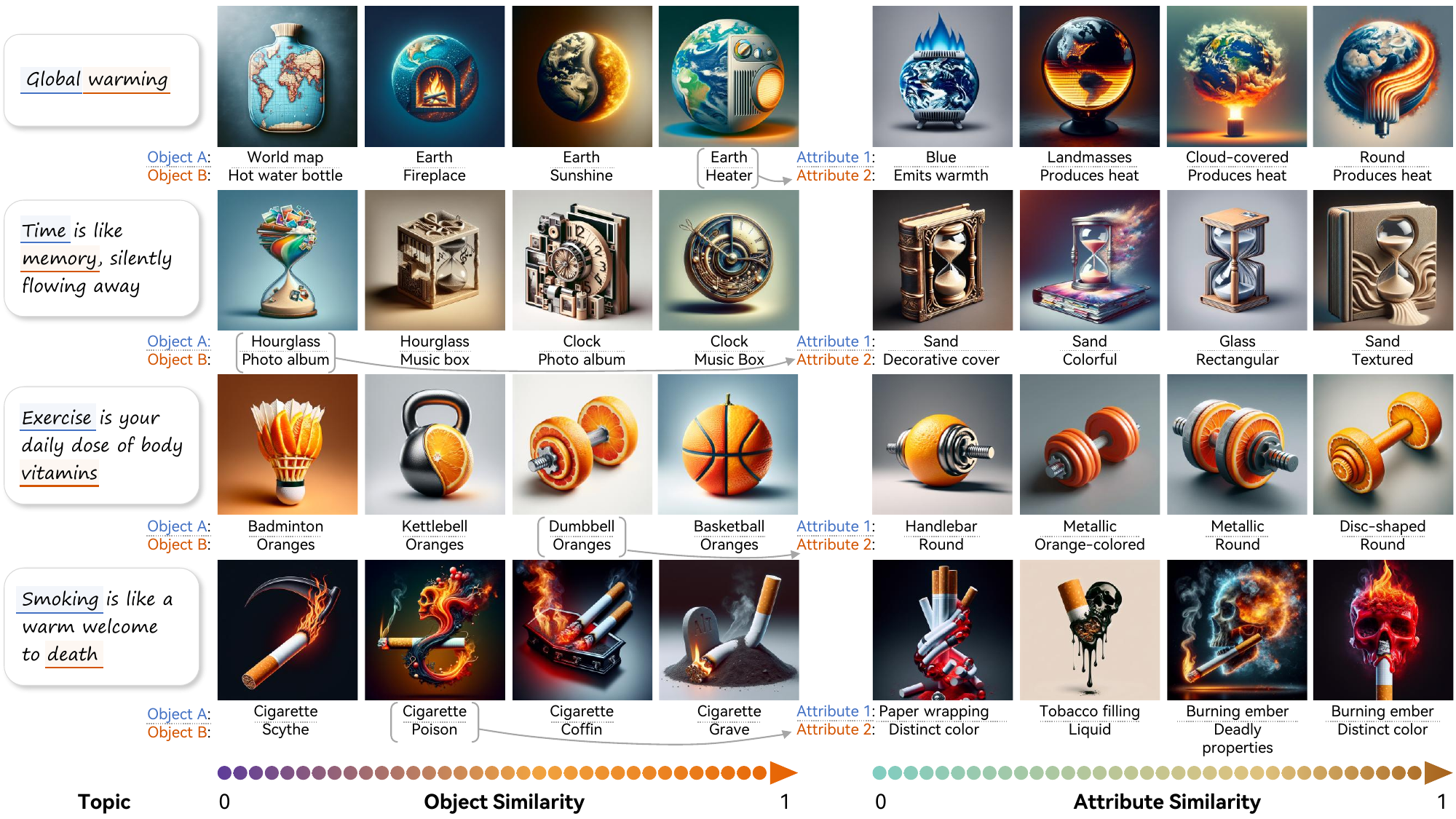}
  \caption{\sysname{} generates diverse visual blends representing abstract concepts based on user-provided expressions. Each topic includes eight examples: four highlighting different levels of object similarity and four demonstrating varying attribute similarity. Similarity increases from left to right. The attributes are extended based on the objects enclosed by the double brackets. Colors within the topics serve to identify concepts and their associated objects and attributes.}
  \label{fig:results}
\end{figure*}

\subsubsection{Metaphoricity}

To assess the metaphorical quality of the generated outputs, we adopted a method outlined in previous research~\cite{10.1007/978-0-85729-224-7_13}.
Participants were asked to indicate the extent to which they perceived a metaphorical relationship between the target and source elements, using a 7-point Likert scale ranging from ``no metaphor'' to ``extremely strong metaphor''.
The Shapiro-Wilk test confirms the normality of the distribution ($W = 0.935, p = 0.124$). 
A paired samples t-test reveals a significant difference in overall similarity between \sysname\ and the baseline ($t(23) = -2.276, p = 0.032$).
The moderate effect size (Cohen's $d = 0.558$) indicates a meaningful improvement in the metaphorical quality of content generated by \sysname, leading to the acceptance of hypothesis \textbf{H5}.

Participants' feedback highlights the significance of metaphor in bridging the gap between abstract concepts and familiar imagery.
As P15 (Male, 23) observed, ``\textit{... the (baseline) often turns abstract information into scenes rather than specific characters}''.
This shows that existing models are not yet able to associate abstract meanings with existing imagery of physical objects. 
In addition, six participants noted that the target and source elements within \sysname\ outcomes were more readily recognizable compared to the baseline (P5, Male, 21; P7, Female, 21; P18, Female, 23; P21, Male, 20; P22, Male, 26; P24, Female, 23).
These findings suggest that the \sysname\ system preserves the original features of blended objects while aligning metaphorical relationships with human cognition within the constraints of the commonsense knowledge derived from the physical world.

\subsection{Ideation Results}

We collected the outputs produced by participants during the study.
These results have been documented and are provided as part of our supplementary materials.
Figure~\ref{fig:user-study} presents eight representative examples selected from four participants (i.e., P1, P5, P6, and P18), showcasing typical outcomes generated by both system conditions. 
Our analysis revealed that both conditions are capable of generating images that use visual metaphors to represent abstract concepts.
However, \sysname\ exhibited a notable ability to generate blended results that align with the definition of visual blends (as shown in Figure~\ref{fig:user-study}). 
In contrast, the baseline often relied on background elements, scene composition, or visual juxtaposition to communicate abstract ideas, and, in some cases, simply incorporated related text directly within the images.

To demonstrate the capability of \sysname, we showcase generated results for four additional topics in Figure~\ref{fig:results}.  
These examples illustrate the versatility of \sysname's output.
The examples in the left part of Figure~\ref{fig:results} show the system's ability to explore physical objects that vary in their relevance to abstract concepts.
To achieve optimal object combinations, users can adjust the blending structure and connection method according to the similarity of their attributes (Figure~\ref{fig:results} right).
Overall, \sysname\ facilitates metaphorical mapping between physical objects and their associated abstract concepts, and builds connections between these different objects, stimulating creative ideas for visual blend creation.

\section{Discussion}

This section delves into user preferences for creating visual blends and the generalizability of our method. We also provide insights for designing with generative AI and highlight the study's limitations.

\subsection{Design Preferences of Visual Blends}

During the evaluation of \sysname, we observed a variety of user designs reflecting their individual preferences.
These preferences were apparent not only in our interviews but also in how they interacted with both systems. 
We can categorize these preferences into three main design options: a preference for the visual expression of metaphors, the features prioritized for blending, and the visual style of blended results.

\subsubsection{Explicit vs. Implicit Expression of Visual Metaphors}

\begin{figure}[t]
  \centering
  \includegraphics[width=0.9\columnwidth]{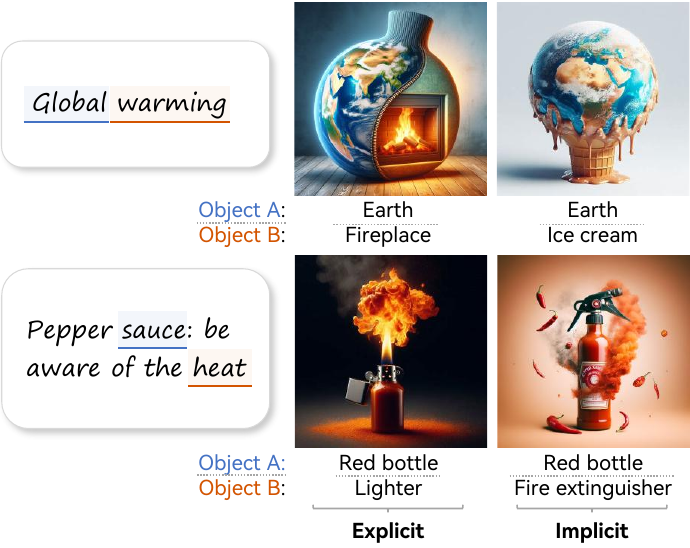}
  \caption{Designers have two options when choosing visual metaphors: explicit (left) or implicit (right).
  }
  \label{fig:direction}
\end{figure}

The design content of visual blends can directly or indirectly represent the topic.
For instance, in Figure~\ref{fig:direction}, when creating visual blends for ``global warming'', we can choose the ``fireplace'' from the explicit direction to establish a direct metaphorical connection with the concept of ``warming''. Alternatively, we can select the ``ice cream'' from the implicit direction to represent this concept through its melting state. Similarly, for the task ``Pepper sauce: be aware of the heat'', we can opt for the ``lighter'' from the explicit direction to directly represent the concept of ``heat''. Or, we can choose the ``fire extinguisher'' from the implicit direction to represent this concept through its fire-extinguishing or heat-reducing function.

Almost half of the participants (13 out of 24) preferred implicit expression, finding it more subtle (P20, Female, 31; P24, Female, 23) and thought-provoking (P13, Female, 25). This approach allows the audience to gain a deeper understanding of the design output. These advantages contribute to the engagement (P11, Male, 27) and imagination (P9, Male, 21) of the resulting outcomes (P12, Female, 24). P9 (Male, 21) noted that implicit expressions provide richer metaphorical information.
While implicit expressions offer these benefits, explicit visual metaphors succeed in directly representing the literal information, providing a clear and straightforward understanding.

\subsubsection{Prioritizing Semantic vs. Visual Features when Blending}

When visually representing abstract concepts, a debate exists regarding the optimal object features to utilize.
Participants' preferences were evenly distributed, with eight out of 24 participants favoring semantic features, while nine favored visual features.
Those who preferred semantic features believed that semantic meaning captures the contextual nuances of concepts, enabling accurate representation (P20, Female, 31) and facilitating the understanding of abstract concepts and relationships that are not immediately apparent visually (P18, Female, 23). 
They hypothesized that blended results could convey the topic more directly, clearly, and comprehensively (P15, Male, 23).
However, participants also recognized the potential ambiguity of semantic meanings arising from polysemy (words with multiple meanings) and homonymy (different words that share the same spelling).
Conversely, those who preferred visual features favored a straightforward utilization of visual elements such as color, shape, and texture when creating visual blends (P8, Female, 21; P14, Male, 22). 
They believed that higher visual similarity would result in more harmonious blended visual outcomes (P6, Female, 23). 
Nevertheless, the superficiality of visual representation can also lead to misinterpretations of meaning.
We propose that when creating visual blends, designers should seek suitable objects based on semantic features and then identify the most appropriate forms of related objects for blending using visual features to achieve more effective conceptual expression.

Beyond using relevant features to facilitate harmonious blending, similarity scores within each feature also play a pivotal role in guiding designers toward diverse exploration paths. 
While high similarity scores are often desirable, designers may intentionally select object combinations with lower similarity scores to achieve more exaggerated and visually striking results.
As illustrated by the example ``exercise is your daily dose of body vitamins'' in Figure \ref{fig:results}, ``orange'' and ``basketball'' exhibit greater visual compatibility than ``orange'' and ``badminton'' and are easier to combine; however, this may lead to a less innovative visual outcome.
This approach encourages designers to explore a broader range of possibilities, stimulating their creativity and accelerating the ideation process.

\subsubsection{The Integrated vs. Independent Styles of Blended Outcomes}

Upon analyzing the generated visual blends, we identified two primary blending styles: objects merging into a unified form while retaining their individual identities or remaining independent and connected by shared attributes.
For instance, the visual blends of ``global warming'' (first row) in Figure~\ref{fig:results} exhibit a cohesive appearance between different objects.
The integrated representation creates a seamless and unified visual, which can be aesthetically pleasing and easier to interpret as a single entity.
However, achieving a balanced and harmonious blend can be challenging and may necessitate a thorough analysis of the objects' attributes.
In contrast, the visual blends of ``smoking is like a warm welcome to death'' (last row) in Figure~\ref{fig:results} present the distinct characteristics of each object.
Such independent representation allows for more flexibility in design, as objects can be repositioned or modified independently without affecting the overall composition. 
Using common attributes to connect objects can highlight the relationships or interactions among them, providing additional information.
However, the separated objects may not convey a sense of harmony and unity as effectively as the integrated design, and an excessive focus on individual characteristics might detract from the overall message.
Overall, the visual blend's style is largely shaped by the attributes of the combined objects. 
Designers have the flexibility to optimize the representation by substituting associated objects or adjusting visual attributes.

\begin{figure}[b]
  \centering
  \includegraphics[width=0.9\columnwidth]{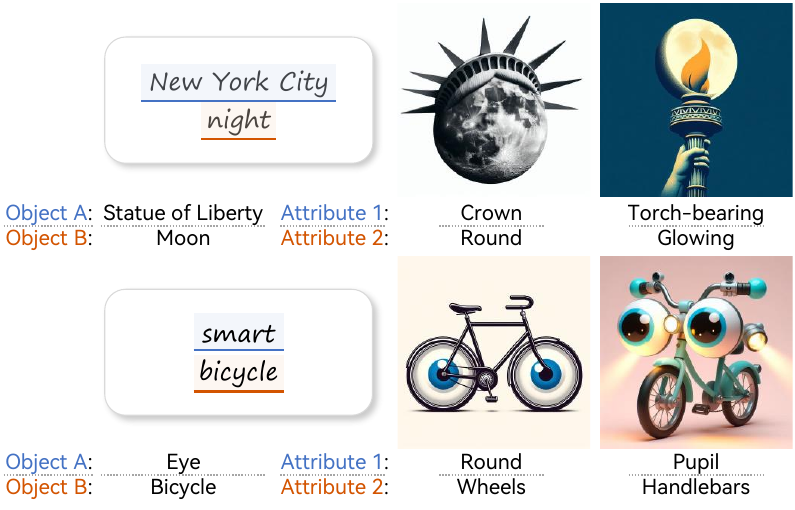}
  \caption{\sysname\ draws upon topics from \textit{VisiBlends}~\cite{10.1145/3290605.3300402} to showcase the related physical objects and the interconnections among their attributes.
  }
  \label{fig:compare-visiblend}
\end{figure}

\subsection{Generalization of \sysname}

\sysname\ is designed and implemented to create visual blends by merging two objects representing distinct concepts.
To test its generalizability, we applied the system to additional topics studied in prior research and expanded our experiments to evaluate its ability to combine multiple objects.

\subsubsection{Create Visual Blends with Previous Topics}

To demonstrate \sysname's performance in generating visual blends, we employed the same topics used in \textit{VisiBlends}~\cite{10.1145/3290605.3300402}.
Unlike the manual brainstorming and image selection processes in \textit{VisiBlends}, our system automatically identifies metaphorically related objects and offers a variety of blending options based on shared attributes. 
This facilitates creative ideation and reduces user effort. 
Figure~\ref{fig:compare-visiblend} demonstrates two topics from \textit{VisiBlends}. 
Beyond the high-quality visual results achieved through advanced image generation techniques, our approach provides a broader range of blending possibilities. 
By considering attribute combinations rather than solely relying on object shapes, we enable more creative blending directions. 
Additionally, our method focuses on the essential aspects of physical objects that connect to abstract concepts, leading to natural and visually appealing blended outcomes.

\begin{figure*}[t]
  \centering
  \includegraphics[width=0.79\linewidth]{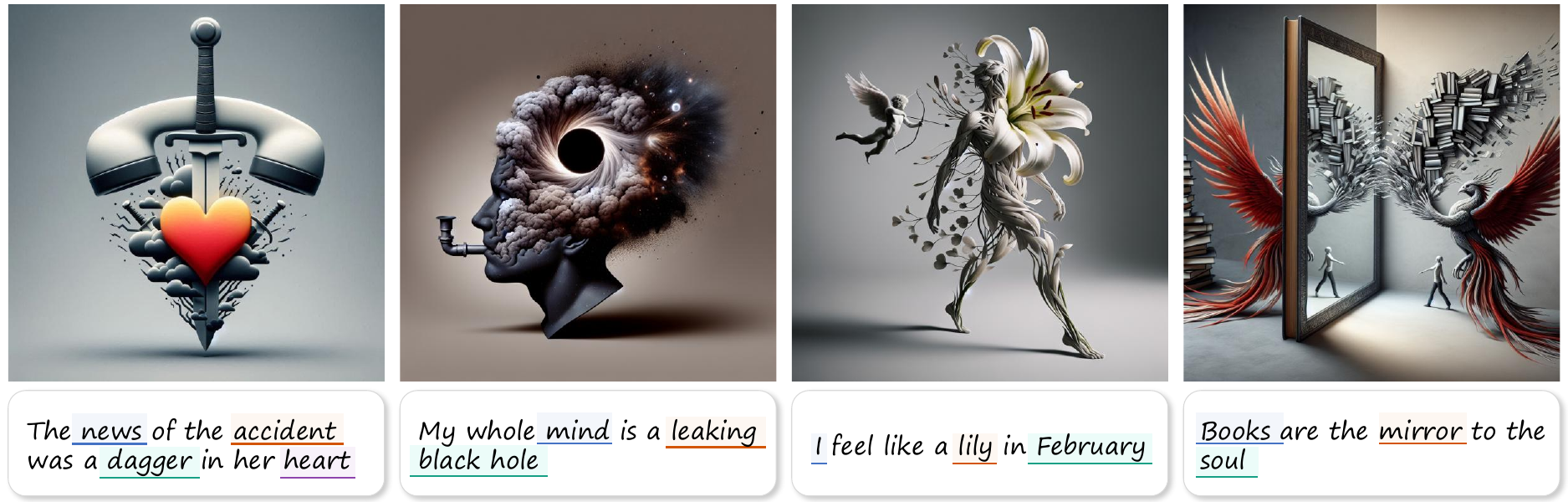}
  \caption{The results generated by our approach when processing more than two concepts. The outcomes presented are in line with the topics explored in \textit{I Spy a Metaphor}~\cite{chakrabarty-etal-2023-spy}.}
  \label{fig:compare-spy}
\end{figure*}

\subsubsection{Increase the Number of Blended Objects}

When blending three or more concepts, our approach emphasizes selecting two main concepts to merge into a primary subject, while the remaining concepts are incorporated sequentially as secondary elements in the image.
By considering similarity and sentiment scores, we strategically select objects and attributes for these additional concepts based on similarity and sentiment scores to create the final blended composition.
Previous work, such as \textit{I Spy a Metaphor}~\cite{chakrabarty-etal-2023-spy}, also explores metaphorical representations.
Unlike its method, which treats all concepts equally, our approach distinguishes itself by emphasizing the blending of specific concepts with metaphorically related objects.
This not only facilitates creative exploration but also enhances the visual representation of abstract ideas~\cite{8eb8812e-2a0d-3ce5-ba0f-6a9667472863}.
Figure~\ref{fig:compare-spy} demonstrates four topics that highlight the differences between our method and the previous approach.
In the expression ``Books are the mirror to the soul'', our system merges ``books'' and ``mirror'' into a primary subject, strengthening the metaphorical connection and establishing it as the main focus of the image. 
Additionally, we incorporate the ``phoenix'' element to represent the ``soul'', showcasing the remaining concepts within the expression.
In terms of the final output, the previous method relied on visual elaboration without clear semantic mapping, leading to scattered visual messaging. 
Our method, however, prioritizes element integration while preserving individual characteristics, making the intended message more easily recognizable by the audience.

\subsection{Design with Generative AI}

During our experiment, we found that the general T2I models do not always reflect users' intent from their generated results. 
Participants commented that ``\textit{sometimes it is tough to keep the AI model on track}'' (P11, Male, 27). 
In some circumstances, users ``\textit{... have no confidence or idea of how the AI will interpret what (they) say}'', and ``\textit{...each prompt is a bit of a gamble}'' (P15, Male, 23). 
These responses revealed two major issues in designing with generative AI.
One is that natural language can be limited in its ability to convey certain design intentions.
The second issue is that the capabilities of generative AI can sometimes be confusing to users due to their unclear boundaries.

The conversational style of current generative AI interfaces has catalyzed a novel trend in interaction design, enabling more intuitive communication with machines through natural language~\cite{10.1007/978-3-031-66329-1_41, 10.1007/978-3-031-48038-6_36}.
However, the lessons we learned have revealed the limitations of natural language in conveying abstract concepts and representing tacit human knowledge, which is often learned through experience rather than explicit instruction.
While natural language interfaces work well in domains where user intent is easily articulated, truly versatile AI interactions require systems that can understand abstract intent and respond to subtle cues, much like human communication.
To mitigate this limitation, specialized tools can tailor the user interface to suit the needs of target users better.
Given the potential shortcomings of natural language interfaces in handling multimedia content~\cite{voigt-etal-2021-challenges}, specialized tools can effectively leverage alternative user interfaces (e.g., graphical user interfaces, or tangible user interfaces) to enhance user support for interactive tasks.

While generative AI offers the potential for processing multi-modal information, it currently requires advancements in task decomposition and domain-specific expertise to address complex problems effectively.
The evolution of software development shows that specialized techniques are needed to harness the unique capabilities of diverse applications. 
Despite the potential to integrate various information sources, current generative AI models sometimes struggle to produce coherent and reliable outputs, frequently generating unrealistic or nonsensical results.
For example, Figure~\ref{fig:user-study} illustrates how P6's result in T1 exhibits a visual discrepancy, such as an improperly lit cigarette. 
While \sysname{} focuses on design ideation, resolving such visual issues falls outside its scope. 
These common failures, often due to training data bias or model misalignment, can be mitigated through techniques such as retrieval-augmented generation or fine-tuning~\cite {liu2024survey}, though these methods require technical expertise and additional resources~\cite{gao2024retrievalaugmentedgenerationlargelanguage}.
From a human-computer interaction perspective, challenges such as ambiguous input, lack of context, and inadequate instructions hinder users' ability to leverage generative AI for complex tasks. 
To improve usability, we propose using prompt engineering to break tasks into smaller sub-tasks and employing a guided task decomposition interface tailored to specific tasks, ultimately enhancing user understanding and problem-solving efficiency.

\subsection{Limitations and Future Work}

Our research is subject to several limitations that we intend to address in future work.
Firstly, our evaluation was conducted using two design ideation tasks, each limited to under 20 minutes. 
In contrast, real-world design processes often span longer durations and encompass multiple iterative rounds of refinement and exploration. 
As a result, our analysis of user interactions and behaviors in creating visual blends was deliberately constrained to the ideation stage, reflecting the system’s primary focus.
In our evaluation, we employed a baseline consisting of Google Search and ChatGPT (GPT-3.5 with DALL·E 3) for comparison, aiming to balance technical sophistication due to the automatic generation capabilities of AI models. 
Other comparisons might involve existing ideation methods like moodboarding~\cite{10.1145/3290605.3300863} or mind-mapping~\cite{10.1145/3411764.3445325} to investigate strengths and weaknesses further.
Future work can also integrate \sysname\ into real-world design cases to observe how the system can facilitate designers' visual blend design processes.

Secondly, our system is designed primarily to inspire users during the ideation process, which currently limits the flexibility of the provided prompts for re-editing.
Moreover, if used to generate the final result, the system still requires multiple trial-and-error iterations to achieve a satisfactory outcome for users. 
Future enhancements could include improving prompt editability to allow professional users greater flexibility in expressing their creativity, incorporating more control modalities to streamline the trial-and-error process, and extending support to the refinement stage further to enhance the system’s utility and creative capabilities.

Thirdly, since our research primarily focused on understanding the system's potential within the design ideation context, a comprehensive analysis of prompt quality and generated result attributes, such as style and layout, was not undertaken.
Nonetheless, we recognize that the quality of user-supplied prompts directly influences the generated outcomes.
Future investigations could delve deeper into how the user's knowledge or experience with AI can affect their experience in generating visual blends. Additionally, exploring the impact of varying prompt formats on AI behavior and the resulting visual blend outputs could provide valuable insights into optimizing the creative process.

\section{Conclusion}

This paper introduces \sysname, an AI-assisted system designed to enhance the process of visual blend ideation by leveraging metaphors. 
Our system utilizes LLMs and commonsense knowledge bases to explore objects and their associated attributes, forming metaphorical connections with abstract concepts. 
It offers the capability to automatically generate blending proposals based on user selections, facilitating rapid creative realization for verification through the T2I model.
To evaluate the system, we conducted a user study involving 24 participants who had AI experience. The findings demonstrate that \sysname\ has the potential to enhance the creativity of the generated ideation results and enable the expression of abstract concepts more metaphorically.
Additionally, this research offers insights into user preferences regarding visual blend design and potential future approaches for supporting design with generative AI.

\section*{Acknowledgments}
We thank all anonymous reviewers for their insightful feedback.
We especially thank Elad Richardson for his valuable input.
This research was supported in parts by ICFCRT (W2441020), Guangdong Basic and Applied Basic Research Foundation (2023B1515120026), Shenzhen Science and Technology Program (KQTD20210811090044003, RCJC20200714114435012, 20231122121504001), NSFC (62472288), ISF (3441/21, 3611/21, 2203/24), and Scientific Development Funds from Shenzhen University.

{\small
\bibliographystyle{ieeenat_fullname}
\bibliography{11_references}
}


\clearpage 
\appendix
\begin{figure}[b]
  \centering
  \includegraphics[width=\linewidth]{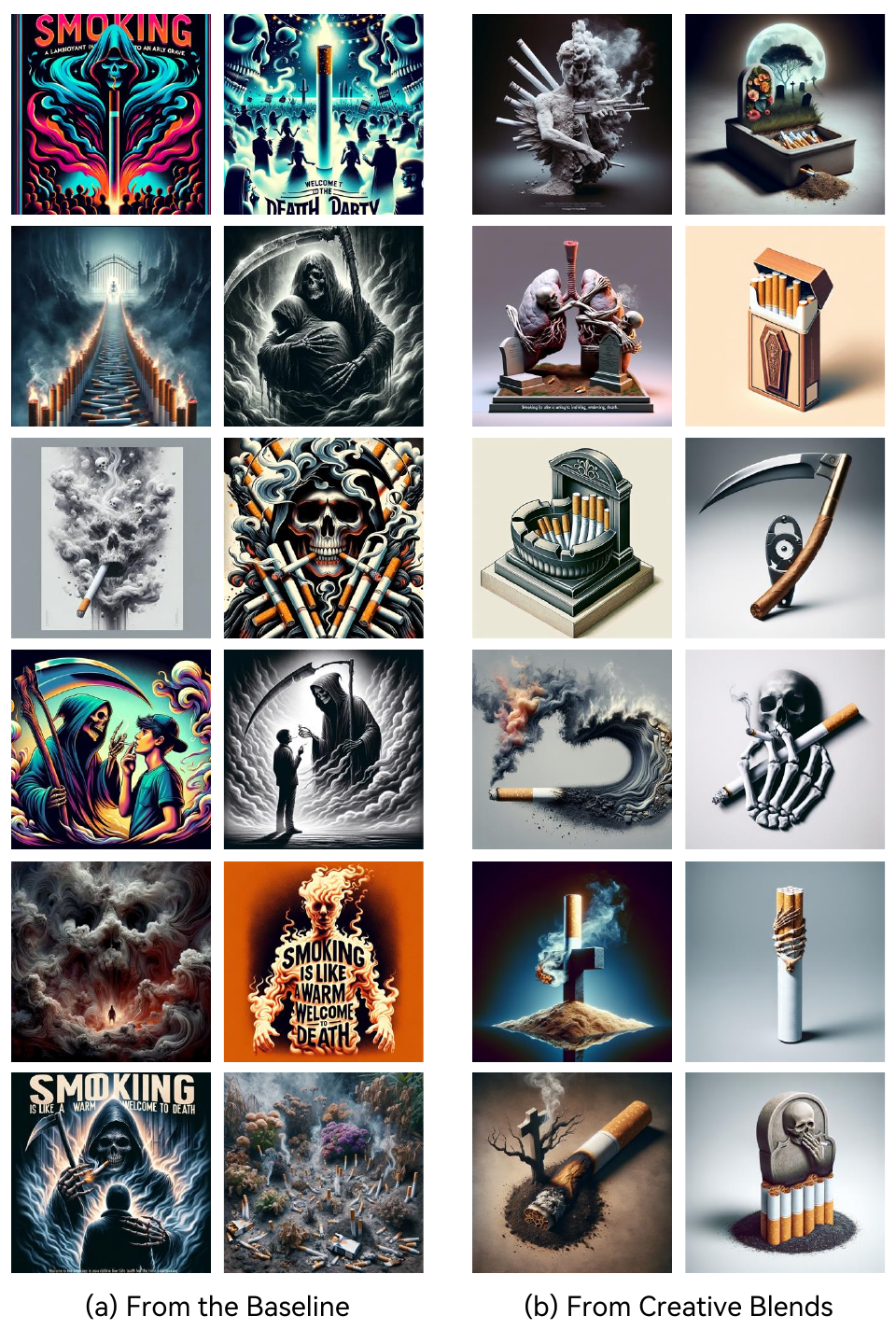}
  \caption{Visual blends for the topic ``smoking is like a warm welcome to death'' (T1).
  }
  \label{fig:study-supp-01}
\end{figure}

\section*{Appendix}
\section{User Study Results}



This section presents the user study results, comparing \sysname\ outcomes to those of a baseline system.
We focus on two specific topics:

\begin{itemize}
    \item ``Smoking is like a warm welcome to death'' (Figure~\ref{fig:study-supp-01})
\end{itemize}

\begin{figure}[b]
  \centering
  \includegraphics[width=\linewidth]{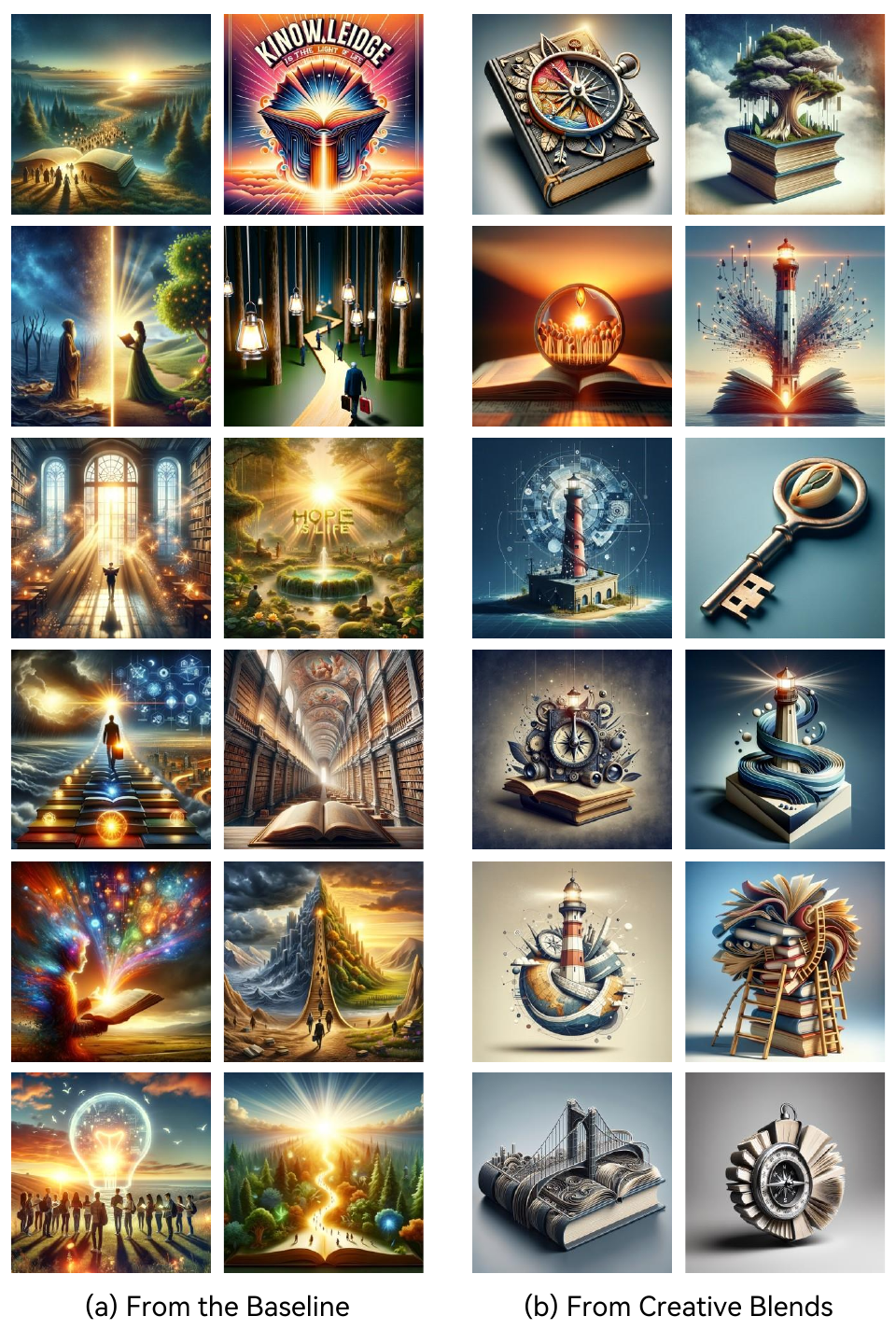}
  \caption{Visual blends for the topic ``knowledge guides the hope of our life'' (T2).
  }
  \label{fig:study-supp-02}
\end{figure}

\begin{itemize}
    \item ``Knowledge guides the hope of our life'' (Figure~\ref{fig:study-supp-02})
\end{itemize}



From the results, it is evident that the baseline system tends to convey the meaning of abstract descriptions through the use of intricate visual elements and by incorporating interaction with human characters.
When encountering abstract content that cannot be directly described with images, text elements are often incorporated within the image itself.
In contrast, the \sysname\ system employs a more imaginative approach, generating visual blends and representing abstract ideas through a variety of objects that are creatively merged to convey metaphorical meaning.




\section{Prompts}

\subsection{Metaphor Generation}

\begin{Verbatim}[breaklines=true]
Prompt = f'''
    system_setup:
        """
        I hope to set you as an assistant with strong reasoning ability and creativity.
        """
    task_definition:
        """
        Our task is to find the hidden meaning in the sentence ({Input}).
        """
    user_input:
        """
        {Input}.
        """
    task_execution:
        """
        Summarize the hidden meaning of the sentence ({Input}) in one sentence.
        """
    result_return:
        """    
        Return the result of your calculation as a valid JSON object. 
        Should be in the following JSON format:
        ###
        {{
        "result": xxx
        }}
        ###
        """
    '''
\end{Verbatim}
\vspace{-8px}

\subsection{Objects Generation}
\begin{Verbatim}[breaklines=true]
Prompt = f'''
system_setup:
    """
    I hope to set you as an assistant with strong reasoning ability and creativity.
    """
task_definition:
    """
    Our task is to support querying a concept, identify multiple physical objects (nouns) associated with it, and reason about the relationships between these physical objects (nouns) and the concept. Then, only return a valid JSON object.
    """
user_input:
    """
    The user input is {INPUT}.
    """
Points to Note:
     """
    1. The content of the answer can not include activities, such as "running" and "swimming".
    2. The content of the answer can not include categories, such as "fruits" and "vegetables".
    3. The content of the answer can not include abstract concepts, such as "exercise" and "beauty".
    """
task_execution:
    """
    1. First, list five specific valid physical objects (nouns), and each of these physical objects (nouns) should have a metaphorical relationship with the {INPUT}.
    The physical object and the {INPUT} can be connected with the "is like" structure, such as "hope is like seeds", and "knowledge is like books". 
    These objects should refer to physical entities (nouns), such as stones and leaves. You can select suitable physical objects (nouns) from external knowledge, or you can reason on your own.
    External knowledge may not necessarily meet the requirements of the task; it requires you to make a selection. 
    External knowledge includes: ###{Related_Concepts}###.
    2. Second, give the reasons why such an object (noun) has a metaphorical relationship with the {INPUT}.
    """
result_return:
    """    
    Return the result of your calculation as a valid JSON object.
    It should be in the following JSON format:
    ###
    {{
    "result": [
    ["physical object(noun) 1", "reason 1"],
    ["physical object(noun) 2", "reason 2"],
    ["physical object(noun) 3", "reason 3"],
    ["physical object(noun) 4", "reason 4"],
    ["physical object(noun) 5", "reason 5"]
    ]
    }}
    ###
    For example, return
    ###
    {{
    "result": [
    ["Dumbbells", "Because they can be used for weightlifting exercises to enhance muscle strength"],
    ["Running Shoes", "Because they are suitable for running exercises to improve cardiovascular function"],
    ["Swimming Goggles", "Because they can protect the eyes and aid in swimming exercises underwater"],
    ["Sportswear", "Because they provide a comfortable wearing experience, making exercise smoother"],
    ["Yoga Mat", "Because they offer a comfortable mat, helping with yoga and balance exercises"]
    ]
    }}
    ###
    """
'''
\end{Verbatim}
\vspace{-10px}

\subsection{Attributes Generation}
\begin{Verbatim}[breaklines=true]
Prompt = f'''
    system_setup:
        """
        I hope to set you as an assistant with strong reasoning ability and creativity.
        """
    task_definition:
        """
        Our task is to receive a list of entities and then show the five most important visible physical attributes of every object.
        Then, only return a valid JSON object.
        """
    user_input:
        """
        The user input is {INPUT}.
        """
    task_execution:
        """
        Please list five attributes for each object. These attributes should be the most important and belong to the physical or tangible properties of the object, such as shape, size, color, or any property that directly describes the object's appearance and structure. 
        Additionally, ensure these attributes are visible to the naked eye, meaning they can be directly observed by human eyes.
        You cannot answer with general terms like "size", "color", or "shape". For example, you should answer with "red" or "round".
        You can select suitable physical attributes from external knowledge, or you can reason on your own.
        External knowledge may not necessarily meet the requirements of the task; it requires you to make a selection.
        External knowledge includes:
        ###{Related_Concepts}###
        """
    result_return:
        """    
        Return the result of your calculation as a valid JSON object. 
        For each object, return [1 object and  5 most important visible physical attributes, total 6 elements].
        It should be in the following JSON format:
        ###
        {{
        "result": [
        ["physical object 1", "<Visible physical attribute 1>", "<Visible physical attribute 2>", "<Visible physical attribute 3>", "<Visible physical attribute 4>", "<Visible physical attribute 5>"],
        ["physical object 2", "<Visible physical attribute 1>", "<Visible physical attribute 2>", "<Visible physical attribute 3>", "<Visible physical attribute 4>", "<Visible physical attribute 5>"],
        ["physical object 3", "<Visible physical attribute 1>", "<Visible physical attribute 2>", "<Visible physical attribute 3>", "<Visible physical attribute 4>", "<Visible physical attribute 5>"],
        ["physical object 4", "<Visible physical attribute 1>", "<Visible physical attribute 2>", "<Visible physical attribute 3>", "<Visible physical attribute 4>", "<Visible physical attribute 5>"],
        ["physical object 5", "<Visible physical attribute 1>", "<Visible physical attribute 2>", "<Visible physical attribute 3>", "<Visible physical attribute 4>", "<Visible physical attribute 5>"]
        ]
        }}
        ###
        For example, return
        ###
        {{
        "result": [
        ["yoga", "stretching", "meditative", "flexible", "focused", "balance"], 
        ["jump rope", "cardiovascular", "coordination", "rhythmic", "agility", "endurance"],
        ["bike", "pedaling", "two-wheeled", "outdoor", "transportation", "gears"], 
        ["hiking", "trail", "scenery", "backpack", "exploration", "nature"], 
        ["hula hoop", "rotating", "waist", "colorful", "spinning", "fun"]]
        }}
        ###
        """
    '''
\end{Verbatim}


\subsection{Scheme Generation}
\begin{Verbatim}[breaklines=true]
Prompt = f'''
system_setup:
    """
    You are a creative assistant for a designer.
    """
task_definition:
    """
    Combine two objects by identifying shared attributes of connection, thereby creating a new object that integrates components from both.
    """
user_input:
    """
    The first object is {Object A}, the shared connecting attribute of {Object A} could be {Attribute 1}.
    The second object is {Object B}, the shared connecting attribute of {Object B} could be {Attribute 2}.
    """
task_execution:
    """
    First, thinking about how to merge {Object A} and {Object B} into one object by utilizing their commonalities: {Attribute 1} of {Object A} and {Attribute 2} of {Object B}.
    Second, justify the rationale for such combination.
    Third, iterate this process to produce (NUM) distinct combinations.
    """
result_return:
    """
    Return the results as a valid JSON object in the following JSON format:
    ###
    {{
    "result":  [
        ["<scheme 1>", "<reason 1>"],
        ["<scheme 2>", "<reason 2>"],
        ...
        ]
    }}
    ###
    For example, return
    ###
    {{
    "result":   [
    ["Merge the shapes of an orange and a dumbbell plate together.", "<Because both the shape of an orange and a dumbbell plate are circular.>"],
    ["<scheme 2>", "<reason 2>"],
    ...
    ]   
    }}
    ###
    """
'''
\end{Verbatim}
\vspace{-10px}

\subsection{Image Generation}
\begin{Verbatim}[breaklines=true]
Prompt = f'''
"Generate an image that creatively blends {Object A} with {Object B}, they should be blended into a single object that has elements from both." + "Highlight the results of blending {Attribute 1} of {Object A} with {Attribute 2} of {Object B} in the resulting blended image." + {selectedScheme} + "The blended image symbolizes a {METAPHORICAL THEME}.The image should have a plain, solid-color background and no text or words."
'''
\end{Verbatim}
















\section{Self-designed Scales for User Study}

To measure outcome satisfaction, aesthetic qualities, and metaphoricity, we use self-developed 7-point Likert scales.
The specific questions are as follows.


\subsection{Outcome Satisfaction}
Higher scores on this measure indicate greater outcome satisfaction.


Q1. To what extent are you satisfied with the \textbf{amount} of outcomes produced?

\begin{figure}[H]
  \centering
  \includegraphics[width=0.85\linewidth]{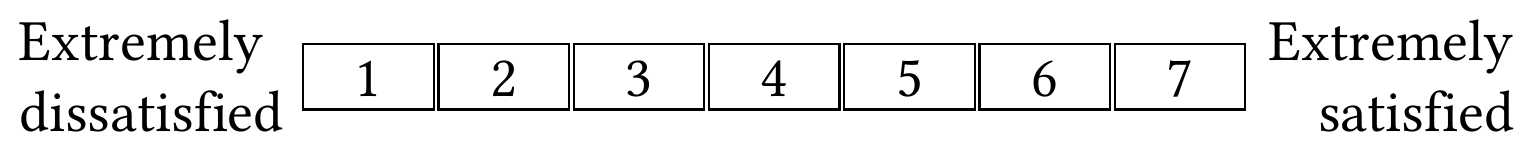}
  \label{fig:supp-01}
\end{figure}

Q2. To what extent are you satisfied with the \textbf{diversity} of outcomes produced?

\begin{figure}[H]
  \centering
  \includegraphics[width=0.85\linewidth]{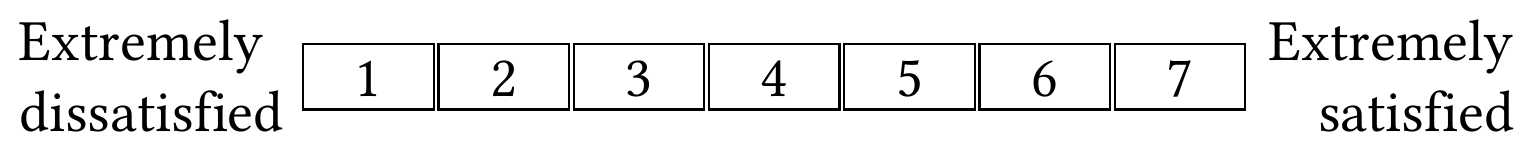}
  \label{fig:supp-02}
\end{figure}



Q3. \textbf{Overall}, to what extent are you satisfied with the outcomes produced?

\begin{figure}[H]
  \centering
  \includegraphics[width=0.85\linewidth]{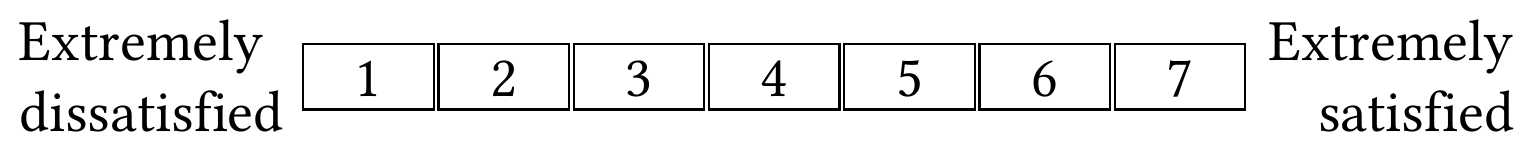}
  \label{fig:supp-03}
\end{figure}

\subsection{Aesthetic Qualities}
Higher scores on this measure indicate greater aesthetic quality.


Q5. How would you rate the overall \textbf{aesthetic quality} of the generated outcome?

\begin{figure}[H]
  \centering
  \includegraphics[width=0.85\linewidth]{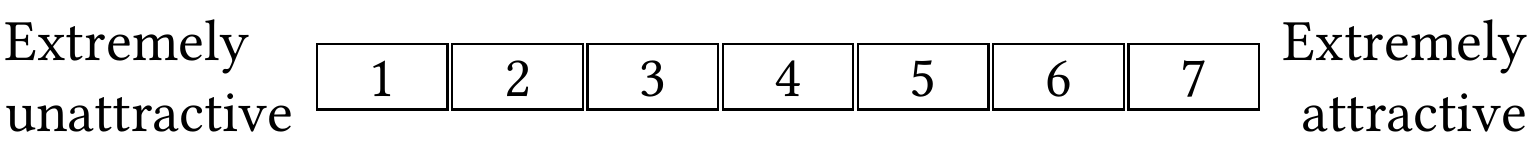}
  \label{fig:supp-04}
\end{figure}

\subsection{Metaphoricity}
Higher scores on this measure indicate greater metaphoricity.

Q6. How would you rate the \textbf{metaphoricity} between the \textit{source} and the \textit{target} in the generated outcomes?

\begin{figure}[H]
  \centering
  \includegraphics[width=0.95\linewidth]{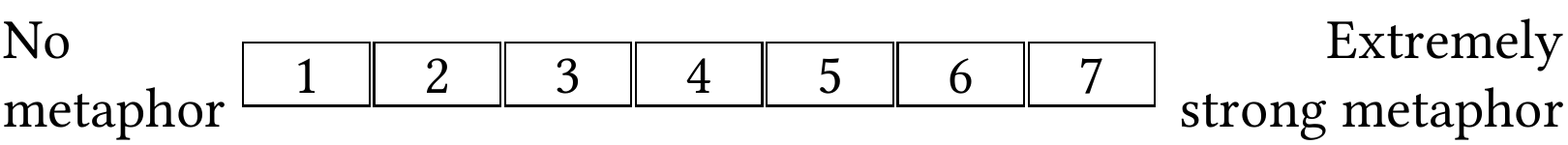}
  \label{fig:supp-05}
\end{figure}

\section{More \sysname: Additional Cases}

\begin{figure*}[h]
  \centering
  \includegraphics[width=\linewidth]{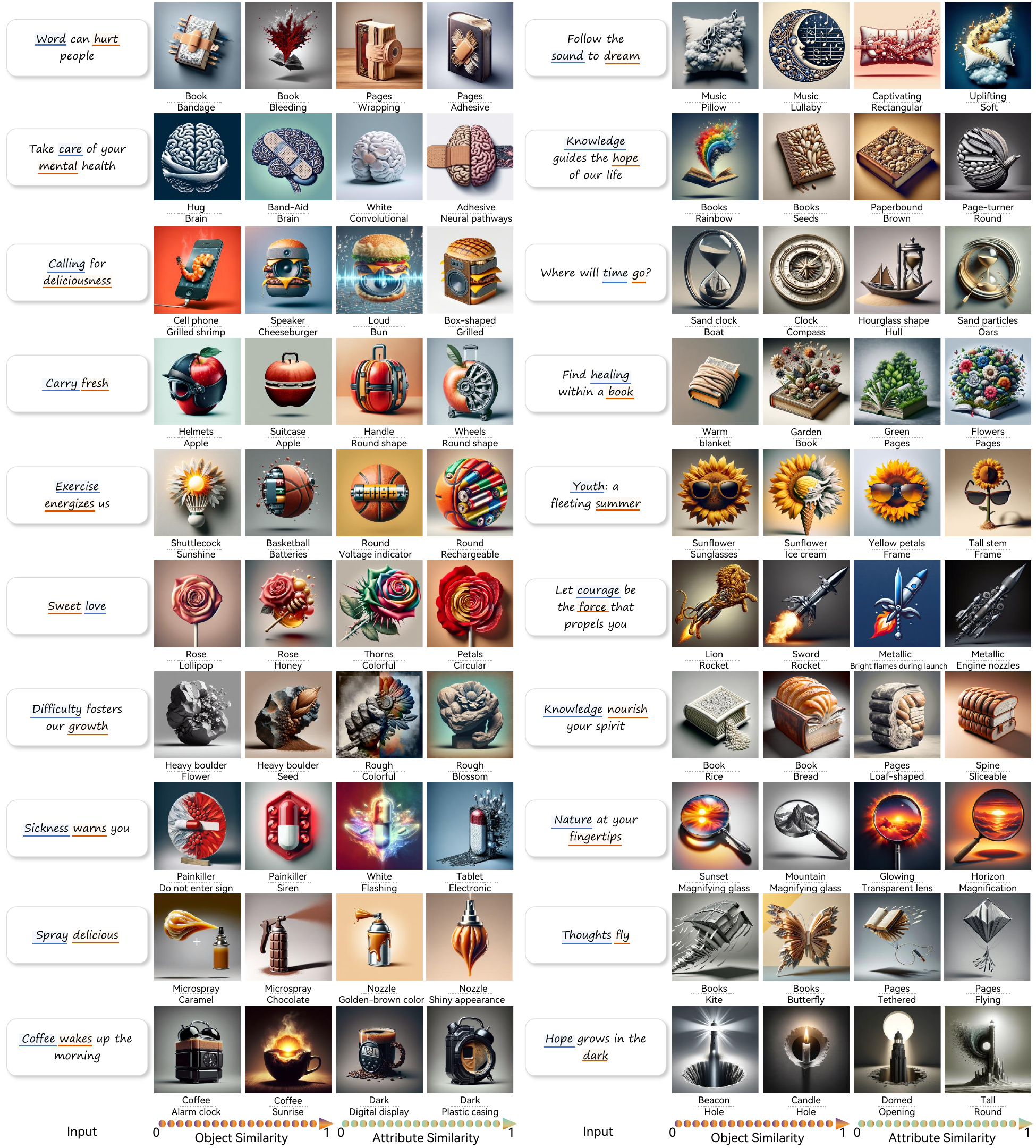}
  \caption{The Sample ideas generated by \sysname. These examples are randomly selected from the topics used in previous research or commonly used in our daily lives.
  }
  \label{fig:supp}
\end{figure*}

In this section, we present a demonstration of the \sysname\ system's adaptability and versatility by showcasing a collection of 20 distinct sets of examples generated by the system (Figure~\ref{fig:supp}).
Through this showcase, we aim to emphasize the system's capacity for generating a wide range of diverse and creative outcomes.













\end{document}